\documentclass[11pt]{article} 

\usepackage[utf8]{inputenc} 
\usepackage[english]{babel}
\usepackage{hyperref}
\usepackage{latexsym}
\usepackage{amsfonts}
\usepackage{times}
\usepackage{xcolor}
\usepackage{graphicx}
\usepackage{xspace}
\usepackage{algorithmicx}
\usepackage{natbib}
\usepackage{authblk}
\usepackage{longtable}
\usepackage{rotating}
\usepackage{pdflscape}

\newcommand*{\affaddr}[1]{#1} 
\newcommand*{\affmark}[1][*]{\textsuperscript{#1}}
\newcommand*{\email}[1]{\texttt{#1}}

\title{\LARGE\textbf{Social Network Analysis:}\protect\\ Bibliographic Network Analysis of the Field and its Evolution \\ Part 1. Basic Statistics and Citation Network Analysis}

\author{
Daria Maltseva\affmark[1] Vladimir Batagelj\affmark[1,2,3]\\
\affaddr{\affmark[1] National Research University Higher School of Economics, Myasnitskaya, 20, 101000 Moscow, Russia.}\\
\affaddr{\affmark[2]Institute of Mathematics, Physics and Mechanics, Jadranska 19, 1000 Ljubljana, Slovenia}\\
\affaddr{\affmark[3]University of Primorska, Andrej Marušič Institute, 6000 Koper, Slovenia}\\ 
\email{d\_malceva@mail.ru}\\
\email{vladimir.batagelj@fmf.uni-lj.si}
}

\newcommand{\keyw}[1]{\textcolor{red}{\emph{#1}}}

\newcommand{\WA}{\mathbf{W\!\!A}}

\newcommand{\WK}{\mathbf{W\!K}}

\newcommand{\WJ}{\mathbf{W\!J}}
\newcommand{\Ci}{\mathbf{Cite}}

\newcommand{\NP}{N\!P}

\newcommand{\Mw}{\mathop{\raisebox{-1.5pt}{\mbox{$\Box$\kern-.55em\raisebox{2.5pt}{{\tiny $r$}}\kern2.9pt}}}}
\newcommand{\Mv}{\mathop{\raisebox{-1.5pt}{\mbox{$\Box$\kern-.55em\raisebox{2.5pt}{{\tiny $h$}}\kern2.9pt}}}}

\newcommand{\Remark}[1]{\ifodd\value{page} \normalmarginpar
 \else \reversemarginpar \fi \marginpar{{\footnotesize #1}} }

\newcommand{\clock}{\count254=\time \divide\count254 by 60
 \count255=\count254 \multiply\count255 by -60
 \advance\count255 by \time
 \ifnum\count254<10 0\fi\number\count254\,:\,%
 \ifnum\count255<10 0\fi\number\count255}

\newcommand{\denseFont}{\fontencoding{T1}\fontfamily{phv}\fontseries{mc}\fontshape{n}\fontsize{10}{11pt}\selectfont}

\graphicspath{{./}{./pics/}}

\begin{document}

\hypersetup{pdfauthor={D. Maltseva, V. Batagelj}}
\hypersetup{pdftitle={SNA. The evolution of the field}}

\maketitle

\begin{abstract}
In this paper, we present the results of the study on the development of social network analysis (SNA) discipline and its evolution over time, using the analysis of bibliographic networks. The dataset consists of articles from the Web of Science Clarivate Analytics database and those published in the main journals in the field (70,000+ publications), created by searching for the key word “social network*.” From the collected data, we constructed several networks (citation and two-mode, linking publications with authors, keywords and journals). Analyzing the obtained networks, we evaluated the trends in the field`s growth, noted the most cited works, created a list of authors and journals with the largest amount of works, and extracted the most often used keywords in the SNA field. Next, using the Search path count approach, we extracted the main path, key-route paths and link islands in the citation network. Based on the probabilistic flow node values, we identified the most important articles. Our results show that authors from the social sciences, who were most active through the whole history of the field development, experienced the ``invasion'' of physicists from 2000's. However, starting from the 2010's, a new very active group  of animal social network analysis has emerged. 
\\[4pt]
\end{abstract}


\section{Introduction}

Social network analysis (SNA) is a rapidly developing scientific field that has appeared and grown significantly over the past 50 years. In the 1970`s the field was highly fragmented and could be represented by a set of individual scientific groups unrelated to each other; these groups existed mostly due to the significant efforts of some individuals and institutions. During 1970-80`s, the \textit{International Network for Social Network Analysis} and \textit{Sunbelt} conference, with specialized journals \textit{Connections} and \textit{Social networks} appeared. In the beginning of 1990`s the representatives of the field have already formed an “invisible college” and the field itself achieved the status of a “normal science” \citep{SNAdev,normSci}. \medskip 

From that point, the field of SNA has grown significantly, both in the number of scientific publications and different disciplines involved \citep{SNAinf, borgatti}. To a large extent, substantial increase in interest in this topic was due to the emergence of the Internet in 1990's and online social networks during the 2000`s. However, if until the 2000`s the field was mostly developed inside different branches of social sciences, starting from the new century it received significant attention from the researchers of the natural science disciplines. The so-called \textit{``  invasion of the physicists'''}  \citep{bonacich} resulted in development of Network Science discipline, whose representatives sometimes were reinventing and rediscovering the issues that had been developed in the social sciences for quite some time \citep{SNAdev}.  \medskip 
 
The development of the SNA field was reflected in a set of studies focused both on its historiographical description \citep{SNAdev,SNAdev2} and bibliometric analysis of publications and journals involved in the field. Several authors studied citation structures of works and journals \citep{normSci,leydes,Understand}, collaboration and co-authorship structures \citep{SNAinf, leydes,Understand}, structures of co-citations between works, authors, and journals \citep{brandes}, topical structures and keyword co-occurence networks \citep{leydes,lookingglass}. Attention was also given to different subfields (subtopics) of SNA \citep{central,kejzar, Understand, batagelj2019} and subdisciplines within the field \citep{SNAinf,borgatti,lazer,varga}. These works are presented in greater details in the following section. \medskip 

In general, various tools of bibliometric analysis has been proposed and extensively used to study scientific disciplines and their development over the last decades. These studies may involve research of various aspects of scientific fields` state and development in different disciplinary and regional areas, such as co-authorship trends in sociology in the USA \citep{moody, sociol}, Slovenia \citep{mali}, or Russia \citep{sokolov}; library and information science in Argentina \citep{rodriguez}, economics in Poland \citep{polish}, the field of  scientometrics and informetrics \citep{hou}, or comparison of different disciplines: biology, physics and mathematics \citep{newman1, newman4}, mathematics and neuroscience \citep{Evol}, or even all research disciplines in a country \citep{kroneg,ferligoj,cugmas}. There are also studies of scientific networks in multinational \citep{glaenzel} and international \citep{wagner} levels. The data for analysis are usually obtained from particular journals \citep{conflict}, thematic sets of literature \citep{dna,PeerRew}, or the databases of bibliographic information \citep{kroneg}. \medskip 

The aim of our study, in line with previous research done in the area, is to implement a comprehensive approach for the identification of the main  trends of the SNA field development, with a representation of various disciplinary areas, groups of scientists, and thematic agenda in the field. The applied bibliometric analysis has already shown to be productive in a set of studies of different scientific fields and topics \citep{kejzar,Understand,PeerRew}. It allows analyzing networks of co-authorship, co-occurrence, citation and co-citation between different bibliographic entities, and identifying key publications and actors (authors, research groups, institutions, journals) in the field of SNA, main topics and scientific ideas, connections between them and their evolution in time. The study is based on the analysis of networks of articles from the \textit{Web of Science} data base and works published in the main journals in the field. \medskip 

Due to the large volume of obtained information, we had to split our results into three parts published separately: (1) basic statistics and citation network analysis (this paper), (2) analysis of co-occurrence networks, and (3) temporal network analysis. The first section of this paper presents some previous studies in the SNA field. Next, we describe the dataset and some issues of network construction. Section four provides some statistical properties of basic networks, and Section five shows the analysis of citation network.

\section{Social network analysis: the review of previous studies}

One of the most comprehensive overviews of the history of SNA development was presented by Freeman in his well-known book \textit{`The Development of Social Network Analysis'}\citep{SNAdev}. Using a methodological perspective of sociology of science, Freeman patterned the links among the people who were involved into the development of the field,  pointed out the main historical events, and thus presented \textit{“the history of social network analysis written from a social network perspective”}. This qualitative study was also supported by the survey of early social network analysts (\textit{`founding fathers'}) on the topic of their introductions to structural thinking -- the scientific antecedents -- and their most important works. \medskip 

According to the history written by Freeman, the birth of the social network thinking can be attributed to the beginning of the 20th century. However, the first more or less consistent period that can be delineated refers to 1940-60`s, which is associated with the emergence of a large number of “schools,” most of which were not aware of each other and were potentially competing. That is why, \textbf{by the 1970`s the field was highly fragmented}: according to the results of the “founding fathers” survey, the field`s intellectual antecedents formed different groups -- sociologists, on the one side (though, loosely connected to each other) and anthropologists, geographers, social psychologists, communication scientists, political scientists, historians and mathematicians (who showed more agreement about the patterns of influence) -- on the other side. \medskip 

Starting from 1970`s, a number of attempts were made for the unification of many separate strands of SNA by a number of individuals and institutions. Among these attempts Freeman points out the organization of the \textit{International Network for Social Network Analysis (INSNA)} in 1977, creation of \textit{Social Networks} journal in 1979, the conferences and the regular meetings that brought separate groups together (including those connected by early version of Internet), the appearance of computer programs standardizing analysis of social network data, educational programs at the universities and “bridging” positions of some scholars travelling around different institutions. All these attempts lead to the \textbf{institutionalization of the field in 1980`s}, when \textit{`the representatives of each of these network “schools” have all joined together and organized themselves into a single coherent field'} \citep[p.~135]{SNAdev}. Freeman also mentions some challenges which the newly established field was facing in the beginning of the 20`th century -- the confrontation between the traditional social network analysts and the physicists, discovering the network approach and \textit{`reinventing existing tools and rediscovering established empirical results'}. \medskip 

The findings of Freeman on the unification of the field are supported by the results of one of the first quantitative study on the SNA field development conducted by \cite{normSci}, which was based on the citation analysis of the works published in the first 12 volumes of \textit{Social Networks} journal and important articles that were cited by their authors. Adding some historiographic data to the results of  network analysis, the authors came to the conclusion that \textbf{by the 1990`s the members of SNA community have met the requirements for being an invisible college}. This notion means that until that time there has been a core active group of scientists (INSNA members), having shared paradigm (understanding of the society as a network), defining important problems, promoting common methods of analysis, and establishing criteria of accomplishment and advance, working in core substantive areas and incrementally developing the ideas. They had primary professional outlet (\textit{Social Networks}) and regular face-to-face interaction (through the conferences). The main paths going through the citation network were few in number, densely connected, extensive in the number of articles linked together, and continuous. That is why Hummon and Carley made a conclusion that the \textbf{SNA not only acceded the status of a discipline, but also that the type of science engaged in within social networks field was what Kuhn had labeled a “normal science”}.\medskip 

Based on the analysis of the number of works related to the SNA field in databases of sociological, psychological and biomedical publications in period 1974-1999, \cite{SNAinf} came to the conclusion that \textit{`it was only in the early 1980`s that SNA started its career'}. Interestingly, while \textbf{the fast growth of number of publications without any sign of decline was mostly seen in the sociology}, the biomedical and psychological literature showed the modest increase as well, which \textit{`proves that other fields, besides Sociology, have used the term and the techniques'} of SNA. Using the information from the \textit{Sociological Abstracts} database, authors also constructed the co-authorship network and extracted the most prolific authors.\medskip 

These `pioneer' works were followed by a number of other studies of the field of SNA and its subtopics and subdisciplines, which used different data analysis methods. Based on the same resource -- \textit{Social Networks} journal -- Leydesdorf, Schank, Scharnhorst, and De Nooy \citeyearpar{leydes} presented the temporal analysis of keywords co-occurrence and co-authorship networks, constructed out of the works published in the period 1988-2007, and extracted the most central figures, belonging to certain branches of the field, and common and specialized topics appearing in the journal`s articles through time. Studying the journal`s citation structures (in both cited and citing dimensions),  authors found its \textbf{strong connection with other sociological journals}, and lower strength connections with journals from psychology, organization and management studies. They also showed that in some years the journal was also cited in a larger citation environment, including journals in physics and applied mathematics. However, \textit{`in spite of the fact that the citation impact of Social Networks in recent years has increased, this has not changed its disciplinary identity'}: it still \textit{`can be considered as a representative of sociology journals'}, rather then an \textit{`interdisciplinary journal'}. In a later study,  Groenewegen, Hellsten, and Leydesdorff \citeyearpar{lookingglass} also combined social network and semantic network analyses to study the developments of content coverage of \textit{Social Networks} and the internal consistency of its community of authors, and analysis of networks of concepts and authors to understand how the community and their interests has developed from 1978.   \medskip 

A comprehensive studies of the SNA field development were made by Batagelj, Doreian, Ferligoj, Kejžar, and others, who studied the collaboration networks among \textit{social network analysts and contributors to network science}, citations between works, and citations between journals \citep{Understand}, based on the data obtained from different databases of bibliographic information. Using variety of networks, constructed out of diverse bibliometric entities (works, authors, journals, keywords, citations, publication year), the analysis of several branches of SNA field was also done on the topics of centrality measures \citep{Understand}, clustering and classification \citep{kejzar}, and blockmodeling \citep{batagelj2019}. The findings of these studies confirmed the trend of the \textit{`invasion to the field'} from other disciplines: while in the early period the SNA field was developed in different branches of social sciences, \textbf{starting from 2000`s, the key highly cited works in the field belonged to the authors from physics (mostly), computer science, neurosciences, and medicine}. The presence of these disciplines in the SNA topic and collaboration structures of the field became more visible. Detailed description of the physicists' appearance in the field of SNA and their tension with social scientists was shown by \cite{SNAdev2}.\medskip  

Using the dataset \textbf{SN5} \citep{sn5} presented by \citeyearpar{Understand} (\textit{Web of Science} descriptions of articles on social networks till 2007) \cite{brandes} implemented the procedure of bibliographic coupling (based on closeness of nodes according to their citing patterns) to different sets of bibliographic entities -- works, authors and journals. The analysis revealed the same patterns that were observed in previous studies: \textbf{the distinction between different groups of authors -- social network scientists and the representatives of Network science discipline} -- with the latter forming the most cohesive groups according to the similarity of citation patterns both in sets of works and authors. The analysis of journals similarity according to their `citation behavior' supported the previous conclusions \citep{normSci, leydes} that \textbf{the field has its own specialty journal \textit{Social Networks}, which is positioned in the group of sociological journals}.\medskip  

Some authors paid attention to the development of the SNA \textbf{within different disciplines}, which in general follows the same trends. In their review of Network analysis usage in \textit{Management and Organizational research}, Borgatti and Foster \citeyearpar{borgatti} also showed the exponential growth of publications in the field indexed by \textit{Sociological Abstracts} and containing “social network” in the abstract or title in the period of 1970-2000`s. Studying \textit{organizational network studies} by means of bibliographic coupling and citation network analysis, Varga and Nemeslaki \citeyearpar{varga} found the strong connection of this field to economics, management and business science, and sociology. \cite{SNAinf}, being interested in \textit{social information discipline}, found the presence of SNA there as well: some of the most active information science authors also published articles in the journals from SNA field (such as \textit{Scientometrics, JASIS(T), Journal of Classification}). Lazer, Mergel, and Friedman \citeyearpar{lazer} studied the development of the SNA field within \textit{sociology} -- \textit{``which has served as the primary home of social network analysis over the last several decades''}. Looking at the co-citation patterns of papers published in two leading general sociological journals, \textit{the American Sociological Review} and \textit{the American Journal of Sociology} at three time points -- 1990-92, 2000 and 2005, they delineated different `canons' typical for different time points and the associated authors in each. Being especially interested in the impact that works written within physics had on the study of social networks within sociology, they found the \textit{``rapid entry of the physicists into the canon between 2000 and 2005, and a possible centralization of the field around small-world networks related research''}.\medskip  

Thus, the previous studies done in the field of SNA development show that the institutionalization of the field reflected in the rapid increase of the yearly number of articles related to it, which was constantly growing from 1970-80`s. According to Freeman, these data show that the study of social networks is rapidly becoming one of the major areas of social science research  \citep{SNAdev}. On the other hand, even though the initial involvement into the field of SNA was interdisciplinary \citep{normSci}, recently the field had to face  some challenges, with  \textit{physicists` invasion} being one of the most important \citep{lazer,brandes,Understand, SNAdev2}. Based on these previous findings, the current study aims to evaluate the main changes that the field came through its history and to highlight the current trends of its development. \medskip


\section{Data}

\subsection{Data collection and cleaning}

The source of data for our research was \textit{Web of Science (WoS)}, Clarivate Analytics’s multidisciplinary databases of bibliographic information. The data set is composed of two parts. It is based on the  SN5 data collected for the Viszards session at the Sunbelt 2008 \citep{Understand}, and contains all the records obtained for the query  \texttt {"social network*"} and articles from the journal \textit{Social Networks}, till 2007. We additionally searched for the works without full descriptions which were most frequently cited and papers on SNA of around one hundred social networkers. The final version of SN5 contained 193,376 works,  7,950 works with a description,  75,930 authors,  14,651 journals, and  29,267 keywords. The SN5 data were extended  in June 2018 using the same search scheme. Starting from 2007, 576 articles from \textit{Social Networks} journal were added. Additionally, in 2018, all the articles from the networks-related journals presented in WoS were included -- such as \textit{Network Science}, \textit{Social Network Analysis and Mining}, \textit{Journal of Complex  Networks} (total 431 article). Other network-related journals -- such as  \textit{Computational Social Networks}, \textit{Applied Network Science}, \textit{Online Social Networks and Media},  \textit{Journal of Social Structure}, and \textit{Connections} -- were considered, but were not abstracted in the WoS. \medskip

Figure~\ref{wos} presents an example of a record describing an article as obtained from WoS. We had to limit our search to the Web of Science Core Collection because for other databases in WoS the CR fields, which contain citation information, could not be exported. \medskip

\begin{figure}
\renewcommand{\baselinestretch}{0.8}
\scriptsize
\begin{verbatim}
PT J
AU GRANOVET.MS
TI STRENGTH OF WEAK TIES
SO AMERICAN JOURNAL OF SOCIOLOGY
LA English
DT Article
C1 JOHNS HOPKINS UNIV, BALTIMORE, MD 21218 USA.
CR BARNES JA, 1969, SOCIAL NETWORKS URBA
   BECKER MH, 1970, AM SOCIOL REV, V35, P267
   BERSCHEID E, 1969, INTERPERSONAL ATTRAC
   BOISSEVAIN J, 1968, MAN, V3, P542
   BOTT E, 1957, FAMILY SOCIAL NETWOR
NR 61
TC 2156
PU UNIV CHICAGO PRESS
PI CHICAGO
PA 5720 S WOODLAWN AVE, CHICAGO, IL 60637
SN 0002-9602
J9 AMER J SOCIOL
JI Am. J. Sociol.
PY 1973
VL 78
IS 6
BP 1360
EP 1380
PG 21
SC Sociology
GA P7726
UT ISI:A1973P772600003
ER
SK IP
\end{verbatim}
\caption{WoS record}\label{wos}
\end{figure}

The works, which appear only in WoS CR fields as references, do not have a full description in the collected data set, and are called \keyw{terminal} works. As such works can be higly cited and in this sense important, we additionally collected full descriptions for works with high (at least 150) citation frequences using WoS. If a description of a work was not available in WoS, we constructed a corresponding description without CR data, searching for the work in Google Scholar (exported in RIS biblographic format and converted into WoS with a special R function). We also included manual descriptions of important works without the CR field from the dataset BM on blockmodeling \citep{batagelj2019}. We should note that additional influential papers, usually published earlier, could be overlooked by our search queries because they do not use the now established terminology. Finally, our data set included 70,792 WoS records with a complete description.  \medskip

Some comments should be given concerning the choice of the dataset for the current study. Even though for a long time \textit{Web of Science} had a monopoly in the field of scientific work abstraction and evaluation, other sources of bibliometric data appeared -- such as \textit{Scopus, Google Scholar}, special citation resources and scientific social media (\textit{SciFinder, Mendeley, etc.}). Previous comparison of different databases has shown that they vary significantly according to their coverage of certain scientific disciplines, and have their pros and cons. For example, \textit{Google Scholar} is shown as providing broad coverage for most disciplines, while \textit{Scopus} and \textit{WoS} are found out to have less publications and weaker represention of the works in the social sciences and the humanities. At he same time the amount of works for all disciplines,  especially for engineering, was found to be higher in \textit{Scopus}, then in \textit{WoS} \citep{hilbert, harzing, martin}. \textit{WoS} contains mainly publications from the \textit{journals} with certain level of \textit{impact factor}, while \textit{Google Scholar} contains different types of sources, including journals, conference papers, books, theses and reports. This can be important for the representations of those disciplines where the journals are not the only prestige sources for scientific knowledge sharing (but also conference proceedings, reports, etc.), and publications are not the only types of scientific contributions (but also software, data, patents, etc.) \citep{franceschet}. We propose that this can lead to certain underrepresentation of some fields in our dataset, where SNA is developing -- for example, \textit{computer science}. At the same time, an important feature of \textit{WoS} is that it provides coverage back to 1900 with descriptions including \textit{references} (CR field); for other databases, the information on citations is included to the descriptions of publications only from 1970 (\textit{Scopus}), or not included at all (\textit{Google Scholar}) \citep{elsevier,wos}. Together with lower consistency and accuracy of data in \textit{Google Scholar}, it makes the choice of \textit{WoS} most appropriate for the current study. However, it should be noted that the \textbf{results are inevitably relative to the available data}.  

\subsection{Basic networks construction}

Using \textbf{WoS2Pajek 1.5} \citep{wos2pajek}, we transformed our data into a collection of networks: one-mode citation network $\Ci$ on works (from the field CR) and two-mode networks -- the authorship network $\WA$ on works $\times$ authors  (from the field AU),  the journalship network $\WJ$ on  works $\times$ journals  (from the field CR or J9), and the keywordship network $\WK$ on works  $\times$ keywords (from the fields ID, DE or TI). An important property of all these networks is that they share the same first node set -- i.e. the set of works (papers, reports, books, etc.) -- wich means that they are \keyw{linked} and can be easily combined using the network multiplication into new \keyw{derived}  networks \citep{Understand}. The reslults of these networks analysis are presented in a separate paper (Part 2).  \medskip

Works that appear in descriptions can be of two types: those which have full descriptions (\textit{hits}), and those which were only cited (listed in the CR fields, but not contained in the hits). These information was stored in a partition $DC$, where $DC[w] = 1$ if a work $w$ has a WoS description, and $DC[w] = 0$ otherwise. Partition $year$ contains the work`s publication year from the fields PY or CR. This information is essential for the construction of temporal networks analyzed in Part 3. Also the vector $\NP$ was obtained, where $\NP[w] =$ number of pages in a work $w$. \textbf{WoS2Pajek} also builds a CSV file \textit{titles} with main data about \textit{hits} (short name, WoS data file line, first author, title, journal, year), which can be used to list the results. \medskip 

The usual \keyw{ISI name} of a work (its description in the field CR) has the following structure: \smallskip
 
 \texttt {AU {+ ', ' +} PY \texttt{+ ', ' +} SO[:20] \texttt{+ ', V' +} VL\texttt{+ ', P' +} BP}  \smallskip\\
(first author's surname, first letters of his/her name, the year of publication, the title of the journal, its volume and the number of starting page; \texttt{+} denotes concatenation), which results in such descriptions as \smallskip

\texttt{GRANOVETTER M, 1985, AM J SOCIOL, V91, P481}\smallskip\\  (all the elements are in the upper case). As in WoS the same work can have different ISI names, \textbf{WoS2Pajek} supports also \keyw{short names} (similar to the names used in HISTCITE output), which has the following format:\smallskip

 \texttt {LastNm[:8] \texttt{+ '\_' +} FirstNm[0] \texttt{+ '(' +} PY\texttt{+ ')' +} VL \texttt{+ ':' +} BP}. \smallskip\\ For example, for the mentioned work the short name is \texttt{GRANOVET\_M(1985)91:481}. From the last names with prefixes \texttt{VAN}, \texttt{DE}, \ldots the spaces are deleted, and unusual names start with characters \texttt{*} or \texttt{\$}.\medskip 

After all iterations of cleaning (see Appendix A for details), we finally constructed the data set used in this paper. From 70,792 hits (works with full description, $DC=1$) we produced networks with sets of the following sizes: works $|W| = 1,297,133$, authors $|A| = 395,971$, journals $|J| = 69,146$, key words $|K| = 32,409$. We also removed multiple links and loops from the networks and labeled the \keyw{obtained basic} networks \textbf{CiteN}, \textbf{WAn}, \textbf{WJn}, and \textbf{WKn} (Table~\ref{rednet}). The statistical properties of these networks are presented in the Section 4. \medskip  

\subsection{Reduced networks construction}

\begin{table}
\caption{Sizes of basic and reduced networks}\label{rednet}\medskip
\begin{center}
\begin{tabular}{c|r|r|r|r}
	&\# nodes (sum)	& \# nodes 1	&\# nodes 2	& \# arcs \\ \hline		 
CiteN & 1,297,133 & & & 2,753,633\\ 
\textbf{CiteR} & \textbf{70,792} & & & 398,199 \\ \hline
WAn	& 1,693,104	& 1,297,133	& 395,971	& 1,442,240 \\ 	
\textbf{WAr}	& 163,803	& \textbf{70,792}	& \textbf{93,011}	& 215,901 \\ \hline
WKn &  	1,329,542	& 1,297,133	& 32,409	& 1,167,666 \\  
\textbf{WKr}	& 103,201	& \textbf{70,792}	& \textbf{32,409}	& 1,167,666 \\ \hline
WJn & 	1,366,279	& 1,297,133	& 69,146	& 720,044    \\ 	
\textbf{WJr} 	& 79,735	& \textbf{70,792}	& \textbf{8,943}	& 61,741 \\ \hline
\end{tabular}				
\end{center}
\end{table}

As it was already explained, for the cited only  works  $(DC=0)$ only partial descriptions are provided: we have information only about the \textit{first} author, the journal and the publication year, and we have no information on the keywords (as there are no titles in ISI names and cited works). That is why for further analysis we constructed networks, which contain only works with complete description $(DC>0)$. All the link weights in the obtained networks were set to 1. We labeled these \keyw{reduced networks} \textbf{CiteR}, \textbf{WAr}, \textbf{WJr}, and \textbf{WKr}. In obtained networks, the sizes of sets are as follows: works $|W| = 70,792$, authors $|A| = 93,011$, journals $|J| = 8,943$, key words $|K| = 32,409$ (remained the same) (Table~\ref{rednet}). \medskip 

\section{Statistics on basic networks}

\normalsize
\subsection{Distributions on CiteN}

\begin{figure}
\centerline{
\includegraphics[width=0.45\textwidth]{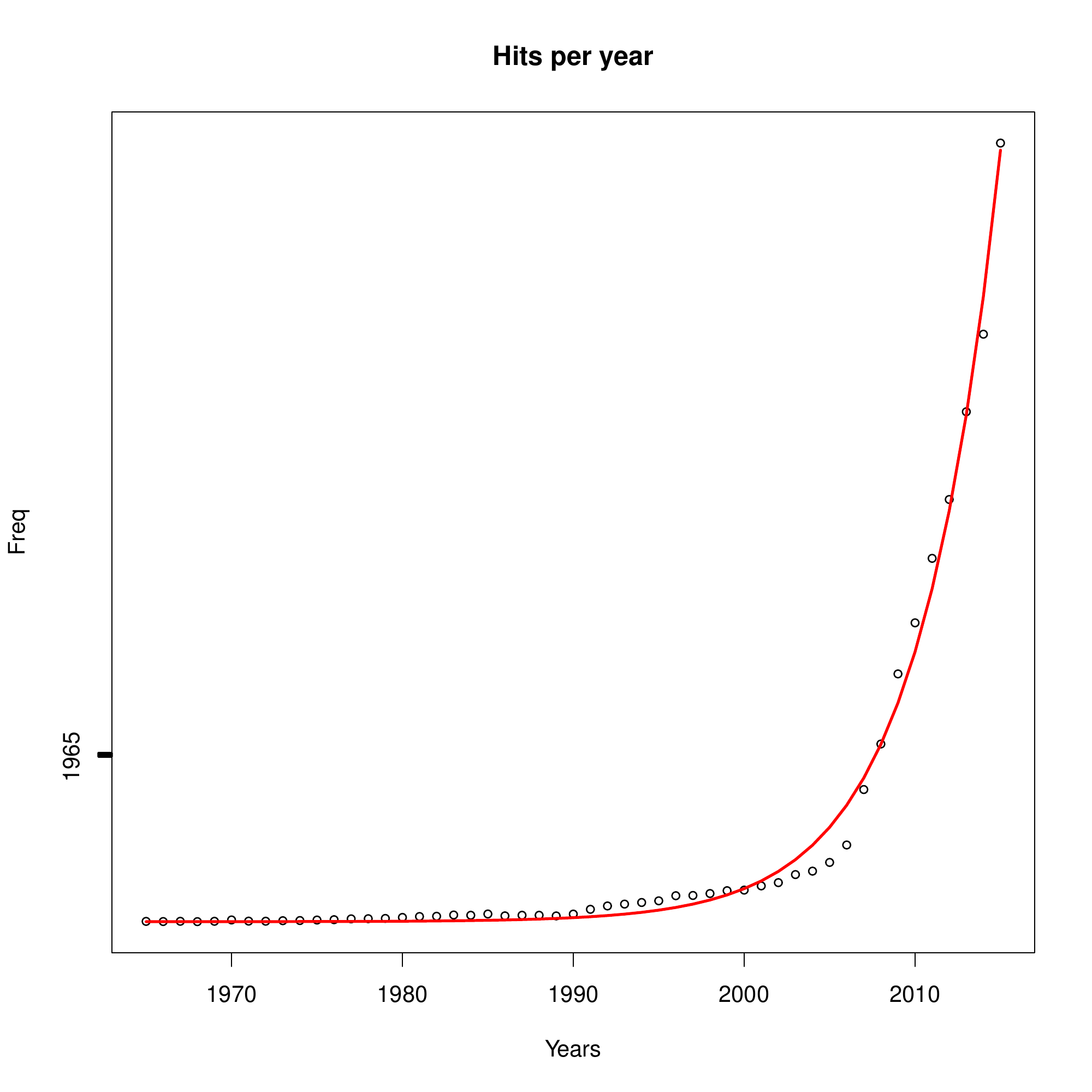} \qquad
\includegraphics[width=0.45\textwidth]{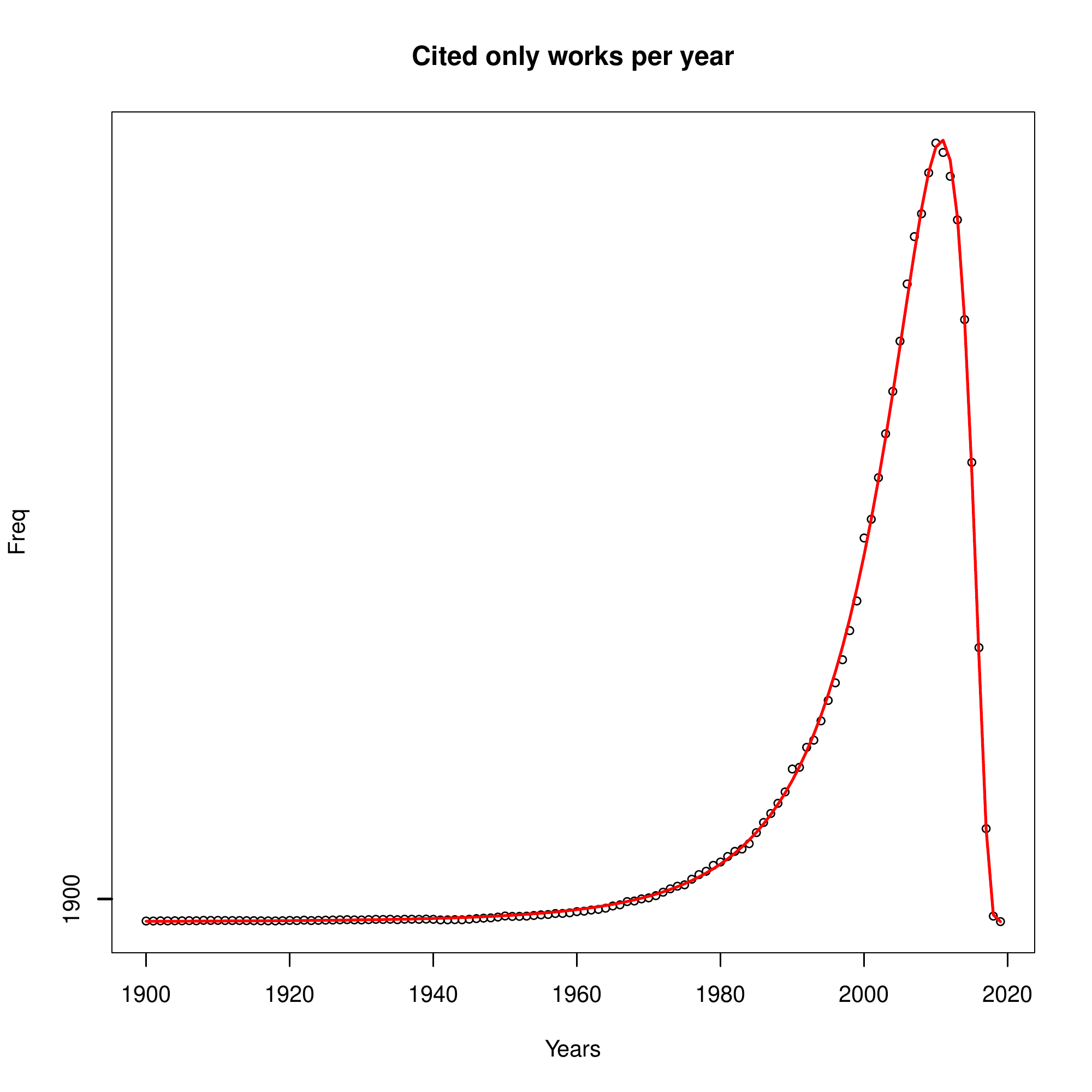} }
\caption{CiteN: Distribution of hits (left) and cited only works (right) by years}\label{yeard}
\end{figure}

In Figure~\ref{yeard}, the distributions of number of works per years are presented. The picture on the left side shows how many works from the set of \textbf{hits} (works with complete description, $DC=1$) are published per year. If the amount of works in our dataset published in 1970 is 21, in 1991 there are already 148 works published, in 2001 - 427, and starting from 2007 the amount of works overcame the level of 1,000: they are 1,576 (2007), 2,119 (2008), 2,955 (2009), 3,564 (2010), 4,333 (2011). Starting from 2012, the amount overcame the level of 5,000: 5,035 (2012), 6,081 (2013), 7,006 (2014), 9,285 (2015), 9,693 (2016). For 2017 and 2018 the amount of works is reduced -- 9,042 and 2,618, respectively -- due to the incompleteness of the WoS data base for recent years. The distribution fits pretty well to the \textbf{exponential model}. The obtained values shows that the amount of works almost doubles in each 3 years ($log(2)/log(1.2338) = 3.299148$). \smallskip 

$c\cdot a^{year - 1965}$, where $a = 1.2338$, and $c = 0.2526$. \medskip

The right side of Figure~\ref{yeard} shows the publication years for the works which are \textbf{cited only} by the hits ($DC=0$). It is clearly seen that the majority of works which are being cited are published recently: if there are 13,202 works published in 1990, starting from 2000 the amount of works is 33,185 (2000), 50,211 (2005), 67,343 (2010). The amounts of works published after 2014 is decreasing: it is 52,074 (2014), 39,724 (2015),  23,704 (2016), 8,045 (2017), and 479 (2018), which simply means that works published recently could not yet get the large amount of citations. However, the presence of the most newest works shows that they are already seen and cited by the representatives of the field. We should also note that there are citations done to the works published in the first part of 20th century and even earlier -- in 14th century (41 works), 15th (20), 16th (45), 17th (245), 18th (528), and 19th (2,151 works). This distribution (from 1900 to 2018) fits very well the \textbf{log normal distribution} \cite[p.~119--121]{Understand}: \smallskip 

$c\cdot \mbox{dlnorm}(2018-year,a,b)$, where $a = 1.501$, $b = 0.9587$, and $c = 7.110\ 10^4$.\medskip

\medskip   

\begin{figure}
\centerline{
\includegraphics[width=0.45\textwidth]{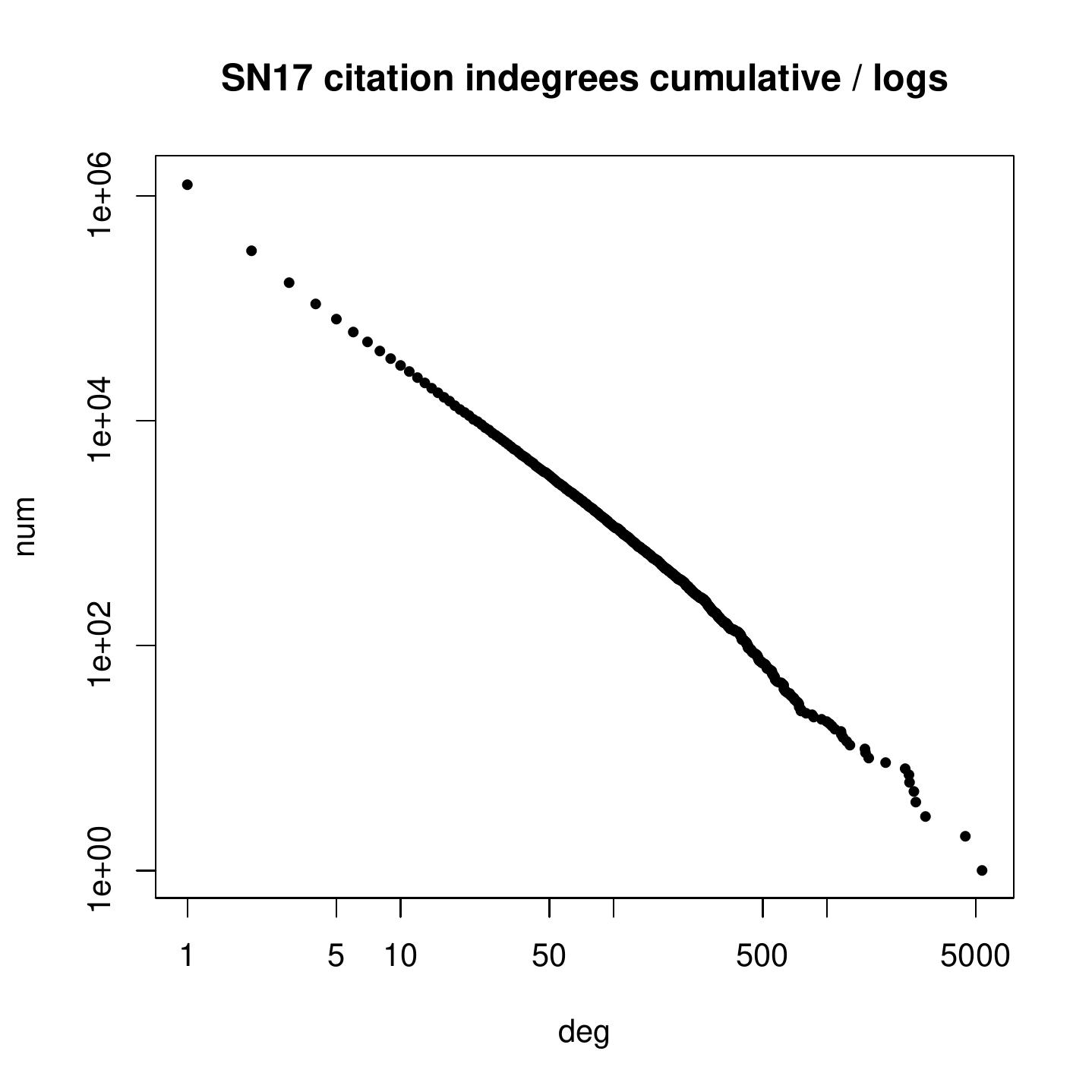} \qquad 
\includegraphics[width=0.45\textwidth]{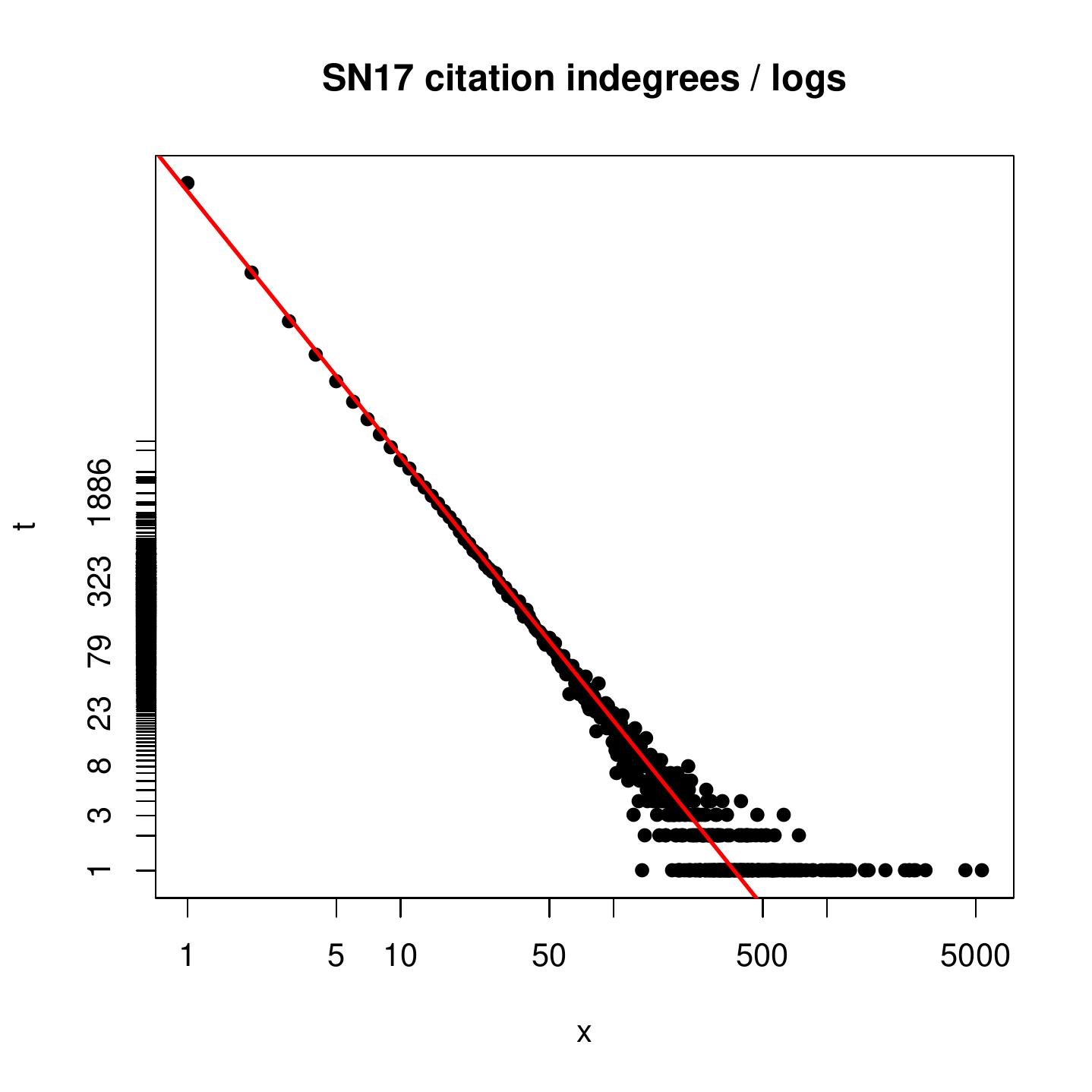} } 
\caption{CiteN: Indegree distribution -- cumulative (left) and density (right) in double-logarithmic scale}\label{cindeg}
\end{figure}

In Figure~\ref{cindeg}, the indegree distribution in \textbf{CiteN} -- cumulative and density -- in double-logarithmic scale is shown. This distribution fits well the \textbf{power law} distribution $f = c \cdot n^{-\alpha}$, with fitted $\alpha = 2.3007$ and $c=749338$. A small number of works attracts a large number of citations, and the large number of works attracts only small number of citations. Works with the largest indegrees are the most cited papers. 

\medskip    

Table~\ref{mostcited} presents 60 most \textit{cited} works (indegree in \textbf{CiteN}). Almost half (28) of these works are published before 2000, and quarter of them (15) are books (their label ends with a colon ``:''). The most cited \textit{book} is the work of \textit{Wasserman and Faust} published in 1994. Other books of networks scientists from the social sciences (marked in boldface, numbers of citations in parentheses) are: \textit{Burt, Structural Holes: The Social Structure of Competition, 1992 (2333); Putnam, Bowling alone: America’s declining social capital, 2000 (1510); Scott, Social Network Analysis: A Handbook, 2000 (1192);  Coleman, Foundations of Social Theory, 1990 (1093); Hanneman, Introduction to social network methods, 2005 (854); Lin, Social capital. A theory of social structure and action, 2001 (800); Rogers, Diffusion of innovations, 2003 (628); Putnam, Making democracy work: Civic institutions in modern Italy, 1993 (613); Zachary, An information flow model for conflict and fission in small groups, 1977 (583); Burt, Brokerage and closure: An introduction to social capital, 2005 (565);  Rogers, Diffusion of Innovation. 4th, 1995 (555);  Fischer, To dwell among friends: Personal networks in town and city, 1982 (539)}. Interestingly, the a book \textit{Ucinet for Windows: Software for Social Network Analysis, 2002} appears twice, attributed to Everett (1171) and Borgatti (999) as the first author. \medskip 

\begin{table}
\begin{center}
\caption{CuteN:\label{mostcited} The most cited works - indegree}\medskip
\renewcommand{\arraystretch}{0.95}
\begin{tabular}{r|r|l||r|r|l}
i	& freq	& id	                                           & i	& freq & id \\ \hline
1& 	5348& 	\textbf{WASSERMA\_S(1994):} & 	31& 	734& 	*NEWMAN\_M(2001)98:404	\\
2& 	4471& 	\textbf{GRANOVET\_M(1973)78:1360} & 	32& 	719& 	*NEWMAN\_M(2010):	\\
3& 	2906& 	*WATTS\_D(1998)393:440& 	33& 	701& 	PORTES\_A(1998)24:1	\\
4& 	2614& 	*BARABASI\_A(1999)286:509& 	34& 	687& 	BLEI\_D(2003)3:993	\\
5& 	2561& 	\textbf{FREEMAN\_L(1979)1:215}& 	35& 	670& 	\textbf{BURT\_R(2004)110:349}	\\
6& 	2447& 	BOYD\_D(2007)13:210& 	36& 	654& 	HANSEN\_M(1999)44:82	\\
7& 	2429& 	\textbf{MCPHERSO\_M(2001)27:415}& 	37& 	639& 	PALLA\_G(2005)435:814	\\
8& 	2330& 	\textbf{BURT\_R(1992):}& 	38& 	634& 	*CLAUSET\_A(2004)70:066111	\\
9& 	1886& 	\textbf{COLEMAN\_J(1988)94:95}& 	39& 	629& 	*BONACICH\_P(1987)92:1170	\\
10& 	1572& 	*NEWMAN\_M(2003)45:167& 	40& 	628& 	ERDOS\_P(1959)6:290	\\
11& 	1520& 	*GIRVAN\_M(2002)99:7821& 	41& 	628& 	\textbf{UZZI\_B(1997)42:35}	\\
12& 	1510& 	\textbf{PUTNAM\_R(2000):}& 	42& 	628& 	\textbf{ROGERS\_E(2003):}	\\
13& 	1285& 	*ALBERT\_R(2002)74:47& 	43& 	613&  \textbf{PUTNAM\_R(1993):}	\\
14& 	1240& 	\textbf{GRANOVET\_M(1985)91:481}& 	44& 	593& 	BERKMAN\_L(1979)109:186	\\
15& 	1192& 	\textbf{SCOTT\_J(2000):}& 	45& 	583& 	\textbf{ZACHARY\_W(1977)33:452}	\\
16& 	1171& 	\textbf{EVERETT\_M(2002):}& 	46& 	572& 	\textbf{BORGATTI\_S(2009)323:892} \\
17& 	1166& 	NEWMAN\_M(2004)69:026113& 	47& 	569& 	*NEWMAN\_M(2001)64:025102	\\
18& 	1093& 	\textbf{COLEMAN\_J(1990):}& 	48& 	565& 	\textbf{BURT\_R(2005):}	\\
19& 	1058& 	STEINFIE\_C(2007)12:1143& 	49& 	561& 	ADLER\_P(2002)27:17	\\
20& 	1034& 	FORTUNAT\_S(2010)486:75& 	50& 	559& 	\textbf{CHRISTAK\_N(2008)358:2249}	\\
21& 	999& 	\textbf{BORGATTI\_S(2002):}& 	51& 	555&  \textbf{ROGERS\_E(1995):}	\\
22& 	945& 	\textbf{CHRISTAK\_N(2007)357:370}& 	52& 	554& 	MILGRAM\_S(1967)1:61	\\
23& 	867& 	\textbf{FREEMAN\_L(1977)40:35}& 	53& 	553& 	BARON\_R(1986)51:1173	\\
24& 	854& 	\textbf{HANNEMAN\_R(2005):}& 	54& 	550& 	\textbf{GRANOVET\_M(1978)83:1420}	\\
25& 	800& 	\textbf{LIN\_N(2001):}& 	55& 	539& 	\textbf{FISCHER\_C(1982):}	\\
26& 	757& 	KAPLAN\_A(2010)53:59& 	56& 	537& 	BRIN\_S(1998)30:107	\\
27& 	756& 	*BLONDEL\_V(2008):P10008& 	57& 	524& 	\textbf{MARSDEN\_P(1990)16:435}	\\
28& 	742& 	NAHAPIET\_J(1998)23:242& 	58& 	523& 	KEMP\_D(2003):137	\\
29& 	740& 	FORNELL\_C(1981)18:39& 	59& 	523& 	KLEINBER\_J(1999)46:604	\\
30& 	740& 	*NEWMAN\_M(2006)103:8577& 	60& 	517& 	*BOCCALET\_S(2006)424:175	\\ \hline
\multicolumn{6}{c}{\textbf{bold} is for social scientists, * for physicists}
\end{tabular}
\end{center}
\end{table}

The second place is taken by a classical article of \textit{Granovetter} on the \textit{strength of weak ties} concept. Other artciles of social network scientists presented in the table are (with topics in parentheses) belong to \textit{McPherson (homophily), Freeman and Bonachich (centrality, betweenness), Burt (structural holes), Coleman, Portes, Adler (social capital), Granovetter, Uzzi (embeddedness), and Milgram (small world)}. \medskip 

The list includes a lot of names of physicists working within the Network approach (marked by *): highly ranked articles of \textit{Watts DJ -- Collective dynamics of 'small-world' networks}, Nature 1998 (2906), as well as \textit{Barabasi AL --  Emergence of scaling in random networks}, Science 1999 (2614). Other works are of \textit{Newman, Albert, Girvan, Fortunato, Blondel, Clauset} on large and complex networks, community detection and clustering. A famous work of mathematicians Erd\H{o}s and Rényi \textit{On random graphs} published in 1959 is also on the list. \medskip 

There are also some representatives of the other disciplines, in  topics such as social network sites and social media (including highly rated article of \textit{Boyd D., Social network sites: Definition, history, and scholarship}, published in 2007 and having 2447 citations); medicine (including well-known works of Christakis and Fowler on spread of obesity and smoking), and management. \medskip 

Works with the largest outdegree in \textbf{CiteN} are the most \emph{citing} works. These works are books, introductory chapters of books, and review articles. Most of these works belong to the field of social sciences, they include education, human relationships, archaeology, migration, internet studies, and social media, but not exactly the topic of SNA. However, some works published in journals in physics and computer science do address the issues of network analysis (Boccaletti on complex networks, Costa on complex networks, Castellano on social physics of social dynamics, Brandes on methodological foundations of network analysis), as well as works representing -- quite surprisingly -- the field of Animal social networks.

\subsection{Distributions on WAr}

As the works with incomplete description (cited only, $DC=0$) contain information only on the first author of works, it is correct to use \textbf{WAr} reduced network to get the information of the number of authors per work and works per author (outdegree and indegree of a network).\medskip

\begin{figure}
\centerline{\includegraphics[width=100mm]{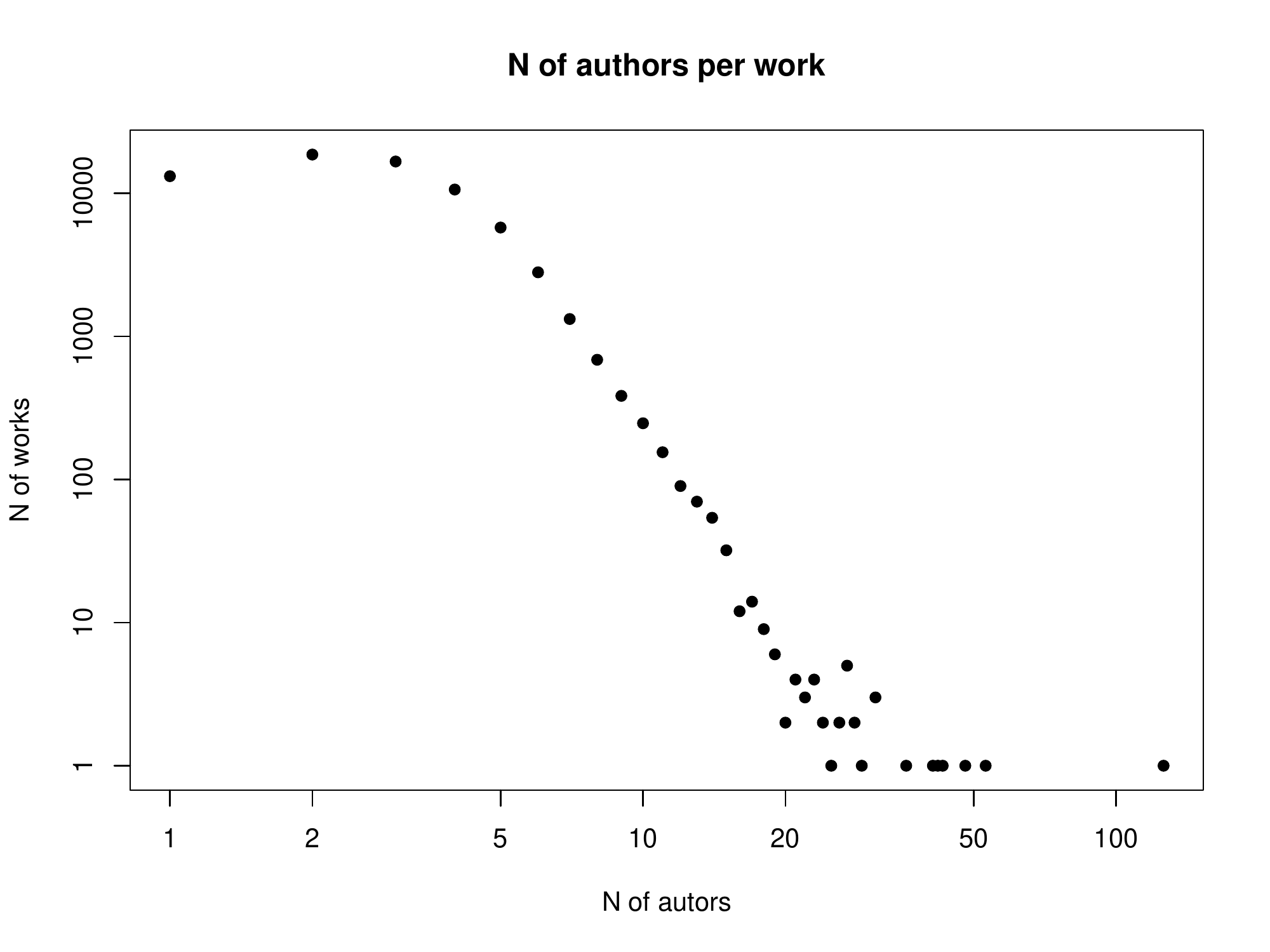}}
\caption{WAr net: Distribution of number of authors by works}\label{authworks}
\end{figure}

The distribution on the number of authors in works according to the reduced \textbf{WAr} network is presented in Figure~\ref{authworks}, and the partition of number of authors in works according to this network in Table~\ref{numpapout}. The majority of works (91\%) are written by 1 author (19\%), or in co-authorship of 2 (26\%), 3 (24\%), 4 (15\%), and 5 (8\%) authors. In some works, however, the amount of authors is pretty high. The extreme' case is the work \textit{Sharing and community curation of mass spectrometry data with Global Natural Products Social Molecular Networking} published in \textit{Nature Biotechnology} in 2016, which has 126 authors. Almost all the works with a large number of co-authors belong to the field of natural science (medical, health, epidemiological, and behavioral studies). For these fields, the inclusion of all authors involved in a research project is quite a frequent practice. However, the third rated article \textit{Discussion on the paper by Handcock, Raftery and Tantrum} published in \textit{Royal Statistical Society Journa Series A: Statistics in Society} collects 48 social networks scientists.\medskip

\medskip   

\begin{table}
\caption{WAr net: \label{numpapout} Number of authors in works -- outdegree} 
\renewcommand{\arraystretch}{0.9}
\begin{center}
\begin{tabular}{l|l|l||l|l|l}
Outdeg&  	Freq&  	Cum freq\% &  	Outdeg&   Freq &	Cum freq\%\\ \hline   
0	&44	&0,062	&19	&6 	&0,008	     \\
1	&13157	&18,585	&20	&2	&0,003	     \\
2	&18635	&26,324	&21	&4	&0,006	     \\
3	&16661	&23,535	&22	&3	&0,004	     \\
4	&10617	&14,997	&23	&4	&0,006	     \\
5	&5759	&8,135	&24	&2	&0,003	     \\
6	&2802	&3,958	&25	&1	&0,001	     \\
7	&1322	&1,867	&26	&2	&0,003	     \\
8	&686	&0,969	&27	&5	&0,007	     \\
9	&384	&0,542	&28	&2	&0,003	     \\
10	&247	&0,349	&29	&1	&0,001	     \\
11	&155	&0,219	&31	&3	&0,004	     \\
12	&90	&0,127	&36	&1	&0,001	     \\
13	&70	&0,099	&41	&1	&0,001	     \\
14	&54	&0,076	&42	&1	&0,001	     \\
15	&32	&0,045	&43	&1	&0,001	     \\
16	&12	&0,017	&48	&1	&0,001	     \\
17	&14	&0,020	&53	&1	&0,001	     \\
18	&9	&0,013	&126	&1	&0,001	     \\		\hline 
Sum	& & & & 70792	& 100,000		\\ \hline	
\end{tabular}
\end{center}
\end{table}

According to the indegree of this network, almost all of the authors with the largest number of papers have Chinese or Korean surnames (Wang, Zhang, Chen, Li, Liu, Lee, Kim, Yang, Wu, Ma). The authors with the largest numbers of works are the following (number of articles in parentheses): WANG\_Y (410), WANG\_X (339), ZHANG\_Y (332), LIU\_Y (321), CHEN\_Y (317), ZHANG\_J (310), LI\_J (305), LI\_Y (304), LI\_X (287). The issue of the super-productivity of these groups of authors was discussed by \cite{harzing2015} -- this is the well-known \href{https://en.wikipedia.org/wiki/List_of_common_Chinese_surnames}{"three Zhang, four Li"} effect: 80\% of people in China have one of only around 100 surnames. Thus, there is a high chance that different authors, having the same surname and first letter of the name, shrink together, creating `multiple personalities'. This problem could be overcame if we would use a special ID (such as ORCID) for each scientist (but this information is not provided in WoS yet). \medskip 

That is why in Table~\ref{numpap} only those authors are presented that did not have Chinese or Korean names. After the serial number, the  number from the original distribution is preserved, so that it can be seen how many authors with ``multiple personalities'' are presented in the data. However, with these names the authors disambiguation problem still occurs, as there are authors with such widespread surnames as Smith, Rodrigues, Johnson, etc. The table list the well-known authors from the SNA field. The most prolific authors are Latkin (130 works), Valente (97), Dunbar (91), Newman (81), Christakis (74), Doreian (72), Carley (72), Burt (71), and others. 

\begin{table}
\caption{WAr net: \label{numpap} Authors with the largest number of papers -- indegree}
\renewcommand{\arraystretch}{0.9}
\begin{center}
\begin{tabular}{|l|l|l|l||l|l|l|l|}  \hline  
Rank & Orig rank & Author & Value & Rank & Orig rank & Author & Value \\ \hline 
1	&45	&LATKIN\_C	&130	    &	   26	&211	&SCHNEIDE\_J	&52   \\
2	&72	&VALENTE\_T	&97	    &	   27	&212	&LEYDESDO\_L	&51     \\
3	&84	&DUNBAR\_R	&91	    &	   28	&217	&LITWIN\_H	&50     \\
4	&102	&NEWMAN\_M	&81	    &	   29	&228	&RICE\_E &48            \\
5	&121	&CHRISTAK\_N	&74	     &	   30	&232	&KAWACHI\_I	&47    \\
6	&126	&DOREIAN\_P	&72	     &	   31	&233	&BONACICH\_P	&46    \\
7	&127	&CARLEY\_K	&72	     &	   32	&234	&PARK\_Y &46           \\
8	&129	&BURT\_R		&71 &	   33	&237	&RODRIGUE\_M  &46      \\
9	&130	&BORGATTI\_S	&71	     &	   34	&238	&NGUYEN\_H	&46    \\
10	&139	&SNIJDERS\_T	&67	     &	   35	&239	&CROFT\_D	&46   \\
11	&140	&BARABASI\_A	&67	     &	   36	&249	&EVERETT\_M	&44     \\
12	&146	&FOWLER\_J	&65	     &	   37	&252	&FERNANDE\_M	&44    \\
13	&149	&KAZIENKO\_P	&64	     &	   38	&255	&CONTI\_M	&44    \\
14	&150	&ROBINS\_G	&64	     &	   39	&256	&MORRIS\_M	&43    \\
15	&152	&WELLMAN\_B	&63	     &	   40	&259	&CONTRACT\_N	&43    \\
16	&163	&FALOUTSO\_C	&60	     &	   41	&266	&WHITE\_H	&42    \\
17	&167	&RAHMAN\_M	&59	     &	   42	&267	&SKVORETZ\_J	&42    \\
18	&172	&PATTISON\_P	&58	     &	   43	&275	&PENTLAND\_A	&41    \\
19	&176	&TUCKER\_J	&58	     &	   44	&276	&WILLIAMS\_M	&41  \\
20	&181	&HOSSAIN\_L	&56	     &	   45	&280	&MOODY\_J	&40  \\
21	&187	&JOHNSON\_J	&54	     &	   46	&289	&FRIEDMAN\_S	&40  \\
22	&194	&NGUYEN\_T	&54	     &	   47	&290	&MARSDEN\_P	&39    \\
23	&196	&MARTINEZ\_M	&53  &	   48	& 292	& BERKMAN\_L	& 39 \\
24	&207	&GONZALEZ\_M	&52  &	   49	  & 301	& KRACKHAR\_D	& 38     \\
25	&209	&RODRIGUE\_J	&52  &	   50	&306	& MORENO\_M	& 38          \\ \hline 
\end{tabular}
\end{center}
\end{table}

\subsection{Distributions on WJn and WJr}

\begin{figure}
\centerline{\includegraphics[width=100mm]{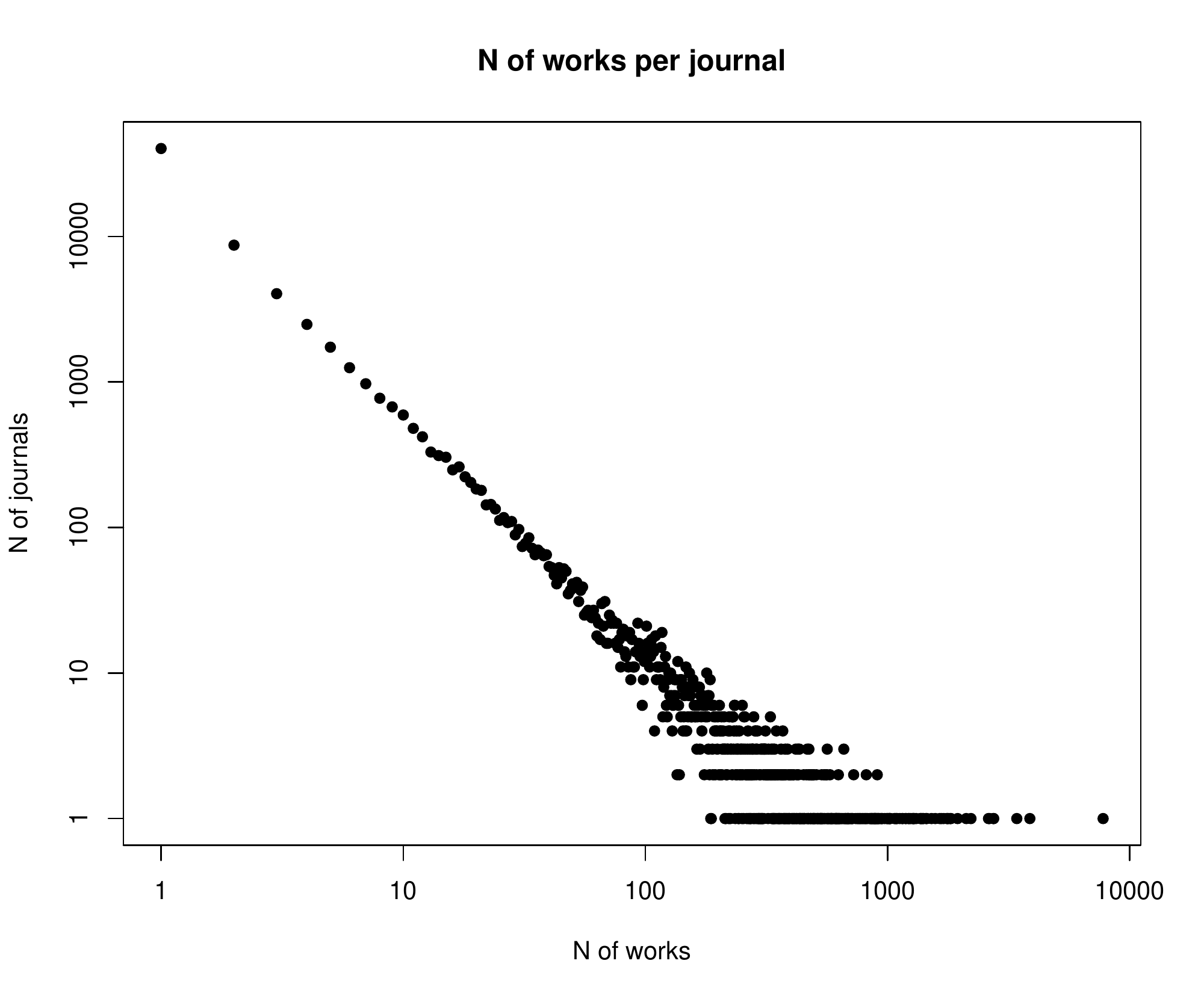}}
\caption{WJn net: Distribution of number of works by journals}\label{worksjour}
\end{figure}

The distribution of number of works per journals is presented in Figure~\ref{worksjour}. It has a scale-free form. According to the indegree distribution of the \textbf{WJn} network, the majority of journals -- in sum, 83\% -- are represented  in the data set with 1 (58\%), 2 (12\%), 3 (6\%), 4 (4\%) or 5 (2.5\%) works. Other 17\% (11,976) journals have 6 works and more. \medskip

Table~\ref{jourind} shows the most used journals, which have the maximum values of the indegree distribution in networks \textbf{WJn} (journals used in all publications) and \textbf{WJr} (journals used in the publications with complete description). The journals in \textit{social sciences} are marked in boldface. \medskip 

The left side of the table presents the indegree from \textbf{WJn} ($DC=0, DC=1$). It contains quite a lot of journals from the social sciences -- such as sociology, psychology, management and business. However, the dominant journal is \textit{Lecture Notes in Computer Science}, which has 7,757 works, followed by \textit{Social Science \& Medicine} and \textit{Journal of Personality and Social Psychology} with more then 3,000 works published. Other journals that have more then 2,000 works are multidisciplinary journals as \textit{Proceedings of the National Academy of Sciences of the USA, Science, Nature}, as well as disciplinary journals \textit{Computers in Human Behavior, American Journal of Public Health, and American Sociological Review}. These journals are followed by other top-ranked journals in different disciplines having more than 1,500 works published, such as (in descending number of works) \textit{Physica A, Animal Behaviour, American Journal of Sociology, Journal of the American Medical Association,  Lancet, Scientometrics, Academy of Management Journal, Lecture Notes in Artificial Intelligence, Journal of Applied Psychology, American Economic Review}. The main field's outlet -- \textit{Social Networks} journal -- is positioned on the18th place, having 1642 works. The remaining journals cover many disciplines such as  Medicine, Psychiatry, Gerontology, Psychology, Management, Marketing, Computer and Information science. \medskip

However, the situation changes quite significantly if we look at the journals with the largest amount of works according to the \textbf{WJr} ($DC=1$) network indegree (right side of the table). The first place is still taken by \textit{Lecture Notes in Computer Science} with 2,009 citations, which is followed by \textit{Social Networks} with 1,134 citations. In this list, the amount of journals from the field of social sciences is less then at the left side. In the WAr network, some journals have shown up in the top, which were not presented in the list of top-40 works in WJn network -- such as \textit{Plos One, Communications in Computer and Information Science, Social Network Analysis and Mining}, and others. Some journals, which were on the top of WJn network indegree distribution, have lowered their positions -- such as \textit{American Journal of Sociology}, -- while some journals have disappeared -- such as \textit{Nature} or \textit{Animal Behaviour}. This means that works from these journals are cited quite intensively by the works found by the network-related search query (hits), but at the same time they are not found by this query, as they have other keywords. Thus, the right side of the table better represents the current thematic directions in the field. 

\medskip   

\begin{table}
\begin{center}
\caption{WJn and WJr nets:\label{jourind} The most used journals -- indegree}\medskip
\small
\renewcommand{\arraystretch}{0.95}
\begin{tabular}{c|r|l||c|r|l} \hline \hline
& \multicolumn{2}{c}{WJn -- Journals used in all publications}		 &  &\multicolumn{2}{c}{WJr -- Journals used by hits}  \\ \hline \hline
 Rank&   	Value&   	Id&   	Rank&   	Value&   	Id \\ \hline
1	&7757	& LECT NOTES COMPUT SC			&  1	&2009	&LECT NOTES COMPUT SC		   \\
2	&3866	& SOC SCI MED				&  2	&1134	&\textbf{*SOC NETWORKS*}			   \\
3	&3414	& \textbf{J PERS SOC PSYCHOL}		&  3	&806	&COMPUT HUM BEHAV		   \\
4	&2741	& P NATL ACAD SCI USA			&  4	&667	&PLOS ONE			   \\
5	&2734	& COMPUT HUM BEHAV			&  5	&531	&LECT NOTES ARTIF INT		   \\
6	&2631	& SCIENCE				&  6	&470	&PHYSICA A			   \\
7	&2609	& AM J PUBLIC HEALTH			&  7	&399	&COMM COM INF SC		   \\
8	&2208	& NATURE				&  8	&375	&SOC SCI MED			   \\
9	&2111	& \textbf{AM SOCIOL REV}		& 9	&319	&\textbf{PROCD SOC BEHV}		   \\
10	&1945	& PHYSICA A				&  10	&314	&PHYS REV E			   \\
11	&1825	& ANIM BEHAV				&  11	&283	&PROCEDIA COMPUTER SCIENCE	   \\
12	&1812	& \textbf{AM J SOCIOL}			&  12	&273	&\textbf{SOC NETW ANAL MIN}		   \\
13	&1780	& JAMA-J AM MED ASSOC			&  13	&238	&ADV INTELL SYST		   \\
14	&1763	& LANCET				&  14	&231	&SCIENTOMETRICS			   \\
15	&1759	& SCIENTOMETRICS			& 15	&225	&CYBERPSYCHOL BEHAV		   \\
16	&1703	& \textbf{ACAD MANAGE J}		& 16	&216	&EDULEARN PROC			   \\
17	&1668	& LECT NOTES ARTIF INT			&  17	&215	&GERONTOLOGIST			   \\
18	&1642	&\textbf{*SOC NETWORKS*}		& 18	&198	&INTED PROC			   \\
19	&1573	&\textbf{J APPL PSYCHOL}		& 19	&194	&SCI REP-UK			   \\
20	&1517	& AM ECON REV				&  20	&188	&J MED INTERNET RES		   \\
21	&1450	& \textbf{J MARRIAGE FAM}		&  21	&186	&P NATL ACAD SCI USA		   \\
22	&1441	& EXPERT SYST APPL			&  22	&180	&EXPERT SYST APPL		   \\
23	&1403	& BRIT MED J				&  23	&176	&INFORM SCI			   \\
24	&1399	& CHILD DEV				&  24	&170	&BMC PUBLIC HEALTH		   \\
25	&1379	& \textbf{RES POLICY}			&  25	&167	&\textbf{NEW MEDIA SOC}			   \\
26	&1372	& COMMUN ACM				&  26	&160	&IEEE T KNOWL DATA EN		   \\
27	&1365	& NEW ENGL J MED			& 27	&153	&IEEE ACCESS			   \\
28	&1311	& PHYS REV E				&  28	&145	&AIDS BEHAV			   \\
29	&1287	& \textbf{SOC FORCES}			&  29	&140	&INFORM COMMUN SOC		   \\
30	&1279	& GERONTOLOGIST				&  30	&139	&STUD COMPUT INTELL		   \\
31	&1278	& \textbf{BRIT J PSYCHIAT}		&  31	&136	&IEEE ICC			   \\
32	&1267	&\textbf{AM J PSYCHIAT}		&  32	&134	&IEEE DATA MINING		   \\
33	&1244	& \textbf{STRATEGIC MANAGE J}		&  33	&132	&\textbf{AM J SOCIOL}			   \\
34	&1225	& \textbf{MANAGE SCI}		&  34	&128	&\textbf{J MATH SOCIOL}	   \\
35	&1221	& \textbf{J BUS RES}		&  35	&120	&IEEE INFOCOM SER		   \\
36	&1189	&\textbf{ACAD MANAGE REV}	&  36	&120	&\textbf{ORGAN SCI}		   \\
37	&1188	&\textbf{J CONSULT CLIN PSYCH}	& 37	&119	&PROC INT CONF DATA		   \\
38	&1154	&\textbf{ORGAN SCI}		&  38	&118	&KNOWL-BASED SYST		   \\
39	&1150	& ADDICTION				&  39	&117	&IFIP ADV INF COMM TE		   \\
40	&1123	& CYBERPSYCHOL BEHAV			&  40	&114	&IEEE GLOB COMM CONF		   \\ \hline 
\end{tabular}
\end{center}
\end{table}

\subsection{Distributions on WKn}

For the works with full description ($DC=1$) the keywords are supposed to be presented in the special fields \texttt {DE} (Author Keywords) and \texttt {ID} (Keywords Plus) of the description. However, for some articles this information is not provided. In such cases the keywords are constructed by \textbf{WoS2Pajek} from the titles of works. All composite keywords were split into single words, and lemmatization was used to deal with the \keyw{word-equivalence problem}. However, the works which are cited only ($DC=0$) do not have keywords. \medskip

The majority of works in \textbf{WKn} (95\%) do not have any keywords - these are the works which do not have a complete description ($DC=0$). The amount of keywords for 70,792 works varies from 1 to 84. The most frequent keywords are presented in Table~\ref{keyind}. Not surprisingly, the words \textit{social} and \textit{network} are mentioned in the largest number of works, followed by \textit{analysis}, which shows the relevance of the data to the topic being studied. Some other frequently used words -- \textit{model, community, graph, structure, relationship, tie} (marked in boldface) -- are related to network analysis, while others - \textit{datum, base, information, research, theory, algorithm, approach, pattern, effect} -- to the scientific research in general. There are also words related to some exact topics which are being studied in network analysis -- \textit{online,  networking, facebook, internet, site, web; health, behavior; support; communication; influence; innovation; trust}. We should note that keywords can have different meanings in different contexts; however, their identification in different subgroups (of authors or works) can give us better understanding of the topic structure of the SNA field. \medskip

\begin{table}
\caption{WKn net: \label{keyind} The most used keywords -- indegree}\medskip
\renewcommand{\arraystretch}{0.9}
\begin{center}
\begin{tabular}{r|r|l||r|r|l}
Rank&  	Value&  	Id&  	Rank&  	Value&  	Id\\ \hline
1&  	51333&  	\textbf{social}&  	31&  	3485&  	\textbf{structure}\\
2&  	46191&  	\textbf{network}&  	32&  	3479&  	life\\
3&  	11751&         \textbf {analysis}&  	33&  	3444&  	risk\\
4&  	10219&  	\textbf{model}&  	34&  	3358&  	research\\
5&  	8104&  	\textbf{community}&  	35&  	3143&  	learn\\
6&  	8090&  	use&  	36&  	3116&  	influence\\
7&  	7596&  	base&  	37&  	3054&  	student\\
8&  	7439&  	information&  	38&  	3054&  	impact\\
9&  	7061&  	health&  	39&  	3049&  	perspective\\
10&  	7023&  	behavior&  	40&  	3042&  	complex\\
11&  	6745&  	online&  	41&  	3024&  	theory\\
12&  	6087&  	networking&  	42&  	2859&  	organization\\
13&  	5833&  	media&  	43&  	2828&  	\textbf{relationship}\\
14&  	5404&  	support&  	44&  	2802&  	algorithm\\
15&  	5101&  	communication&  	45&  	2776&  	education\\
16&  	5013&  	study&  	46&  	2714&  	group\\
17&  	4759&  	datum&  	47&  	2704&  	mobile\\
18&  	4376&  	management&  	48&  	2698&  	\textbf{tie}\\
19&  	4372&  	internet&  	49&  	2695&  	adult\\
20&  	4164&  	knowledge&  	50&  	2633&  	approach\\
21&  	4126&  	user&  	51&  	2608&  	care\\
22&  	4023&  	facebook&  	52&  	2551&  	adolescent\\
23&  	3984&  	technology&  	53&  	2479&  	role\\
24&  	3907&  	site&  	54&  	2472&  	state\\
25&  	3888&  	web&  	55&  	2467&  	innovation\\
26&  	3855&  	self&  	56&  	2434&  	pattern\\
27&  	3784&  	\textbf{graph}&  	57&  	2385&  	effect\\
28&  	3676&  	performance&  	58&  	2339&  	people\\
29&  	3534&  	service&  	59&  	2333&  	trust\\
30&  	3512&  	dynamics&  	60&  	2332&  	family\\ \hline
\end{tabular}
\end{center}

\end{table}

\section{Citation network}  

\subsection{Boundary problem in Citation network}

The original \textbf{CiteN} network  had 1,297,133 nodes. Considering the indegree distribution in this network we got the following  counts for the lowest number of received citations: 0 (41,954), 1 (933,315), 2 (154,895), 3 (58,141), and 4 (29, 885), which all together  cover 94\% of citations. Thus, most of the works were terminal $(DC=0)$ or were referenced only once or twice (indegree = 1 or 2). Therefore, we decided to remove all the `cited only' nodes with indegree smaller then 3 $(DC = 0$ and indeg$<3)$ -- the \keyw{boundary problem} \citep{Understand}. We also removed all the nodes starting with string \texttt{[ANON}.  Finally, we got a subnetwork \textbf{CiteB} with  222,086 nodes and 1,521,434 arcs.

\subsection{Analysis of Citation network}

\begin{figure}
\begin{center}
\includegraphics[width=0.7\textwidth,viewport=95 120 635 606,clip=]{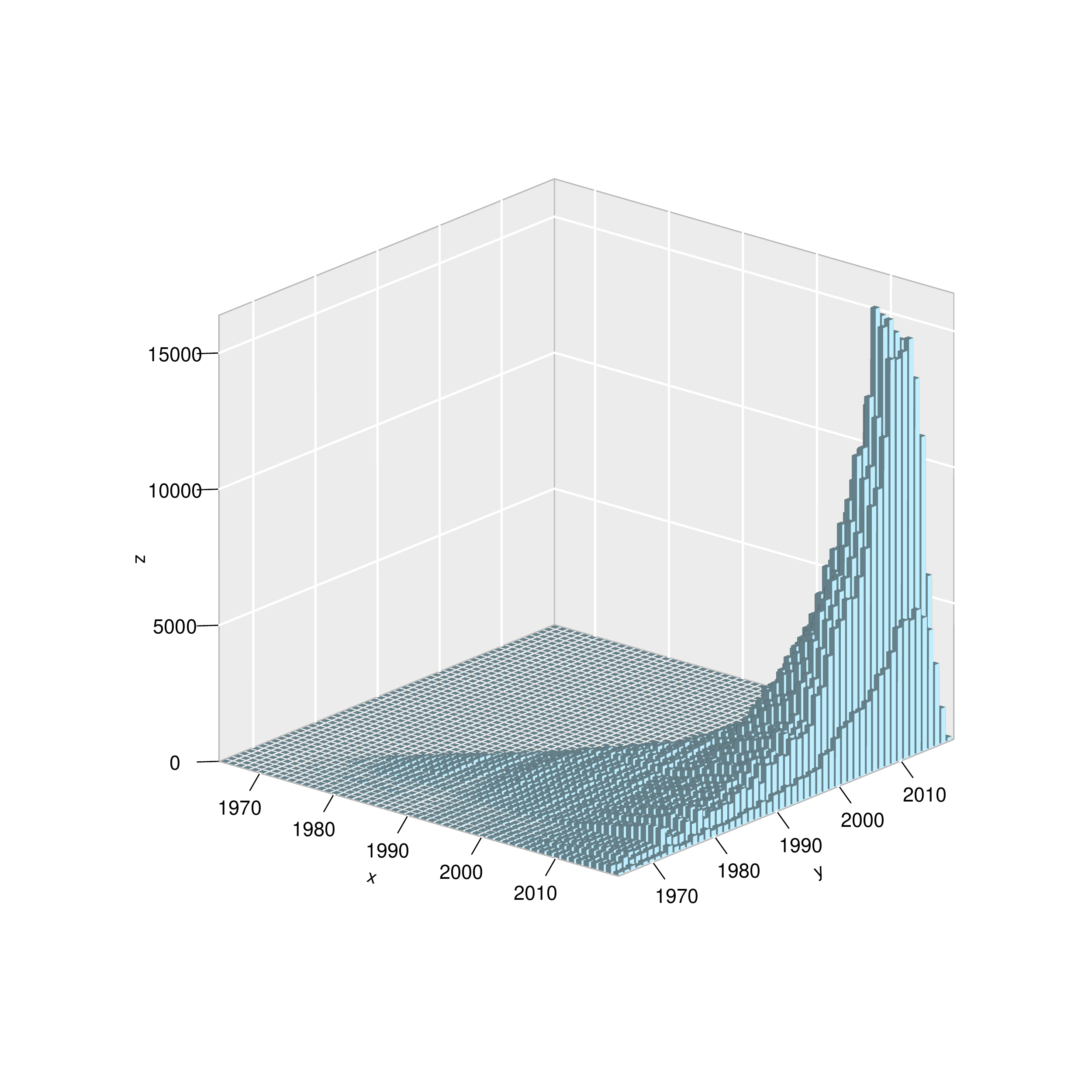}
\caption{Citations between years}\label{yearcite}
\end{center}
\end{figure}

It is interesting to observe how many citations are per years. We combined \textbf{CiteB} network with partition on years \textbf{Years.clu} of publications and constructed the network of  \textbf{citations between years}, where the values of lines are equal to the number of times that all works published in one year were cited in all works published in another year (the network is directed, only later years can cite previous years). Figure~\ref{yearcite} presents the distribution of citations between years in a three-dimensional space. The majority of citations in recent works are done to the recent works as well. The years having the largest amount of citations from other years are 2010 (88,840), 2009 (82,294), 2007 (80,129), 2011 (79,843), 2008 (77,595). Among top-20 years, there are only several years which do not belong to the 2000's: 1999 (39,629), 1998 (36,649), 1997 (27,667), and 1996 (26,216). The largest line weights are from 2015 and 2016 to 2010 (16,384 and 15,755, respectively) and 2011 (16,026 and 15,944). \medskip

\begin{figure}
\begin{center}
\includegraphics[width=0.7\textwidth]{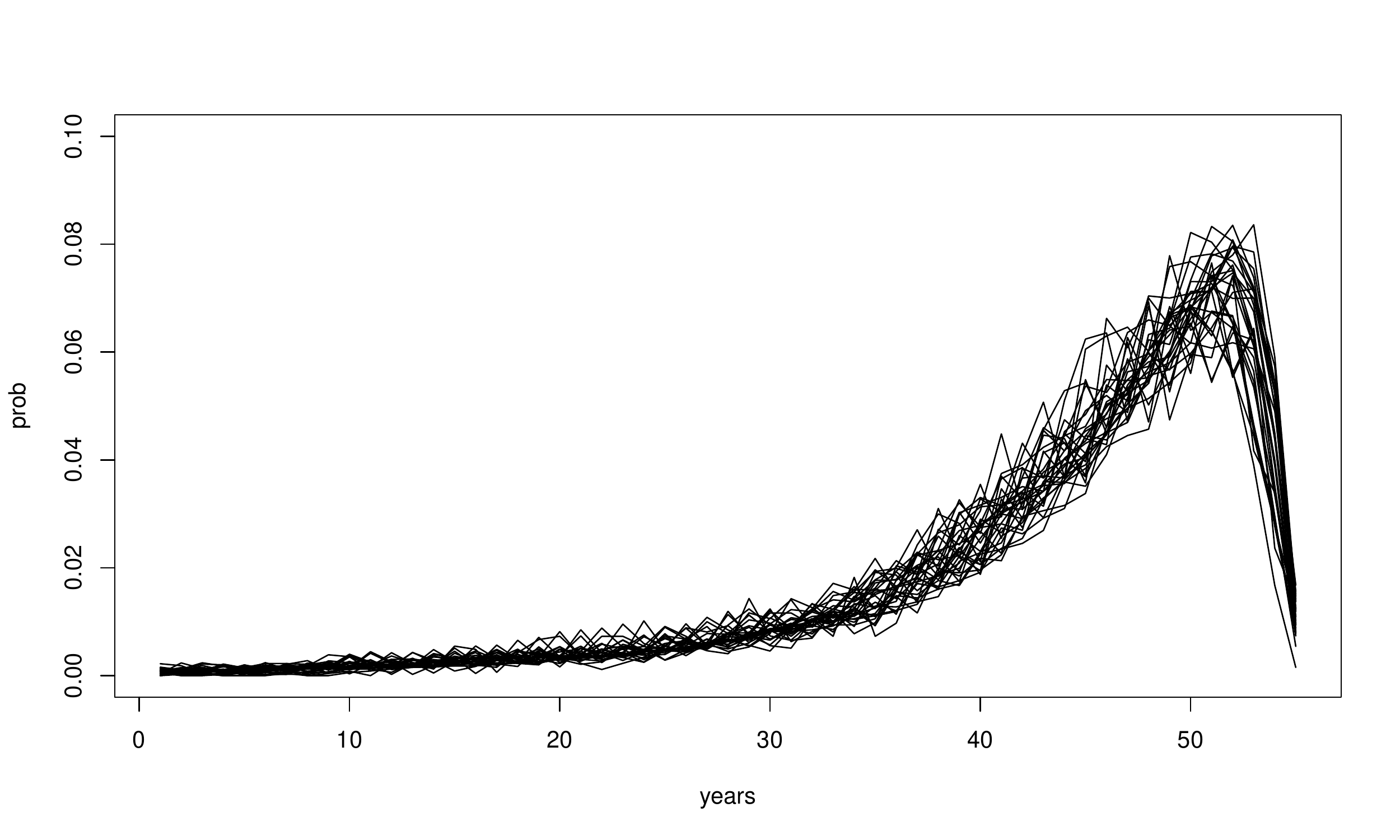}
\caption{Citations between years -- normalized}\label{yearcitenorm}
\end{center}
\end{figure}

Figure~\ref{yearcitenorm} presents the curves of \textbf{normalized values of citations per each year} in the period 1985--2018 (54 years in total). It clearly shows that the yearly citation patterns do not vary significantly from year to year -- there are always noticeably more citations done to the recent works, then to works published previously. This effect was already observed in the analysis of large bibliographic data sets from WoS \citep{lovro}. \medskip

A citation network is usually (almost) acyclic; however, it can include some small cyclic parts, which can be identified as nontrivial strong components of the network (with the minimum size 2). First we searched for nontrivial strong components (see Appendix B for details). To get an acyclic network, as required by the SPC weights algorithm, we applied the \keyw{preprint transformation} to CiteB. The preprint transformation function replaces each work $u$ from a strong component by pair of nodes -- published work $u$ and its preprint version $u'$. Node $u'$ is  labeled by the label of node $u$ preceded by a character ``=''. Published work can cite only preprints. Each strong component was replaced by a corresponding complete bipartite graph on pairs \citep{Understand}. The resulting network \textbf{CiteT} has 222,189 nodes and 1,521,658 arcs.\medskip

We computed the \textbf{SPC weights} on \textbf{CiteT} network arcs \citep{Understand, arxiv}. The normalized SPC weight of an arc is equal to the probability that a random path through the network is passing through this arc. We identified main paths (CPM main path and Key-route paths) in this network, and then used \textbf{Link islands approach} \citep{Understand} to find the most ``important'' parts of this network. 
To find the most ``important'' nodes in the network this approach was supplemented by the \textbf{Node islands approach}; we also computed \textbf{probabilistic flow} for the network  \textbf{CiteT}. 
 
\subsection{CPM main path and Key Routes}  

\begin{figure}
\begin{center}
\includegraphics[width=0.3\textwidth,viewport=118   28 235 262,clip=]{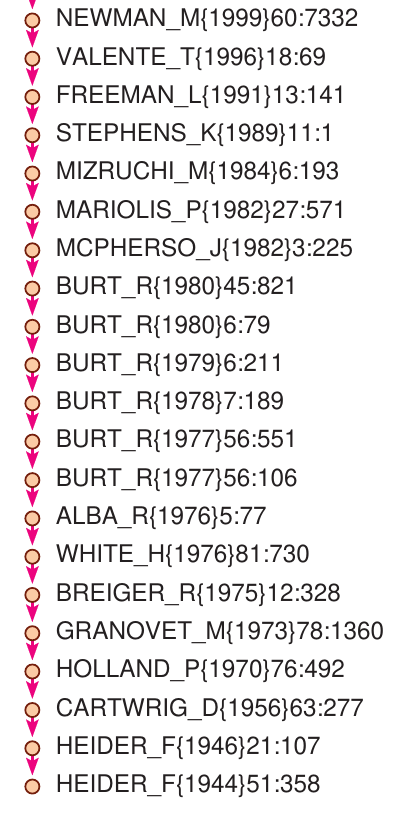}\qquad
\includegraphics[width=0.3\textwidth,viewport=118 239 235 416,clip=]{CPMpath.pdf}\qquad
\includegraphics[width=0.3\textwidth,viewport=118 394 235 681,clip=]{CPMpath.pdf}
\caption{SPC net: Main path by fragments -- sociology, physics, biology}\label{mainFrag} 
\small
(2nd and 3rd parts starts with two works from the previous group)
\end{center}
\end{figure}

Figure~\ref{mainFrag} displays the CPM Main path through the SNA literature (which is the same to the one obtained with the Main path procedure), which includes 59 nodes. We divided this CPM Main path to three parts, according to the disciplinary of the works that are presented. \medskip 

The first group, composed of the works published in 1944 -- 1996, present the works of network scientists from the social sciences. These works appeared (see Appendix C) in journals \textit{Social networks, Administrative Science Quarterly, Annual Review of Sociology, American Sociological Review, Social Forces, Sociological Methods \& Research, Journal of Mathematical Psychology, Psychological Review, The Journal of Psychology}, recalling the history of SNA field formation. In this group, 6 out of 20 works belong to R. Burt. \medskip 

However, since 1999, the initiative in the field goes to the physicists, whose works appear in journals \textit{Physical Review E, Journal of Statistical Physics, Reviews of Modern Physics, European Physical Journal B, Physics Reports, Nature}, and \textit{SIAM Review}. In this part of network, 9 out of 14 works belong to M. Newman. \medskip 

The third part of the Main path, which contains works from 2008 to 2018, is devoted to completely another topic -- animal social networks. The works appear in journals \textit{Animal Behaviour, American Journal of Primatology, Primates, Journal of Evolutionary Biology, Journal of Animal Ecology, Journal of Evolutionary Biology, Trends in Ecology \& Evolution}, and others. The most active author in this group is D. Farine, who has 6 out of 25 works. While the \textit{invasion of physics} into the SNA field was already shown in other studies \citep{lazer,brandes}, the appearance of the third group in the Main path is quite surprising. For the centrality literature analysis it was shown that the trend goes from physics to neuroscience \citep{Understand} and for network clustering literature it consists only of social and physical parts \citep{batagelj2019}  \medskip  

\begin{figure}
\begin{center}
\includegraphics[width=\textwidth,viewport=120 15 605 700,clip=]{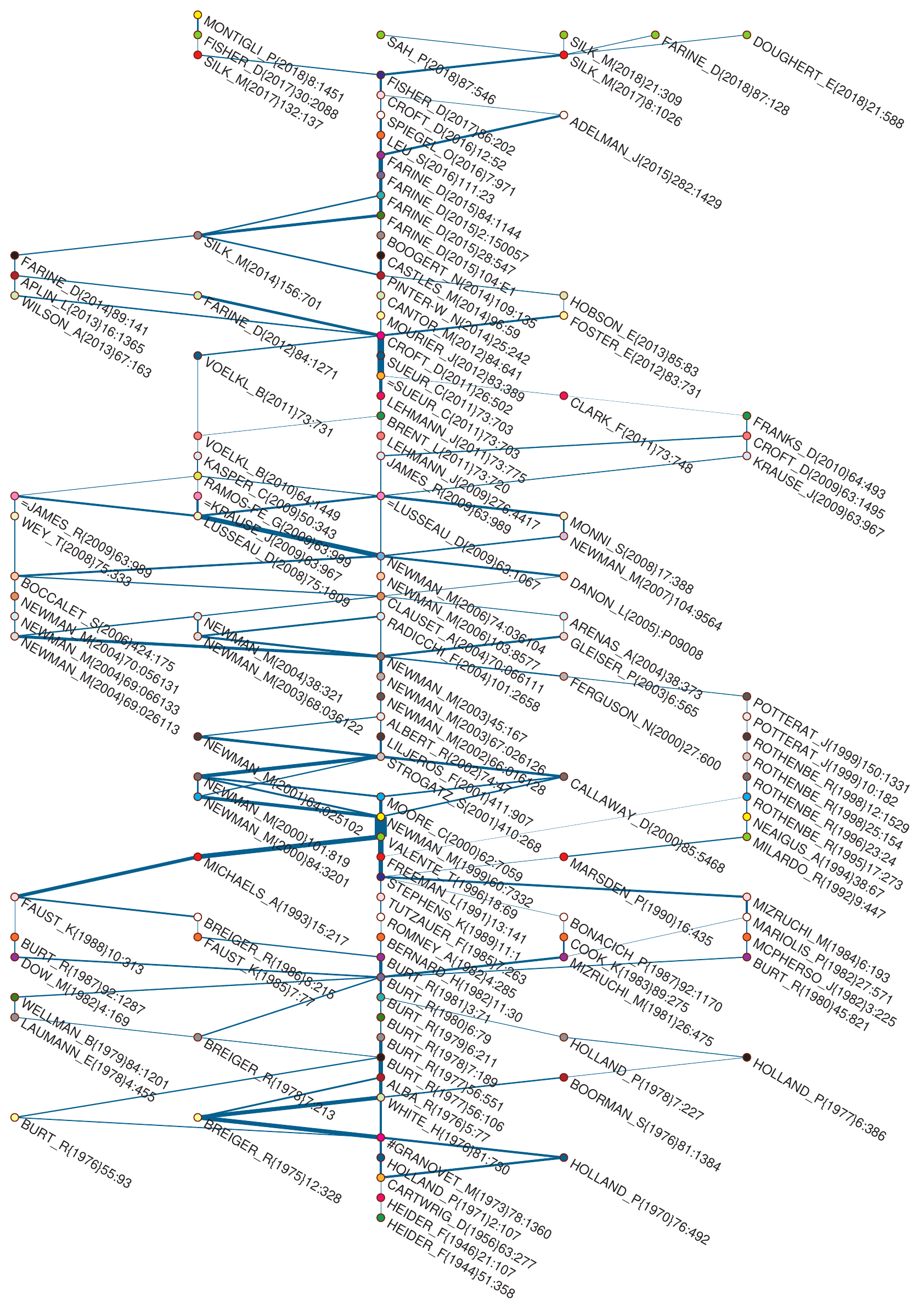}
\end{center}
\caption{SPC net: Key Routes} \label{keyRoute}
\end{figure}

The procedure of Key-route paths \citep{Understand} produces a more nuanced image of most important paths in the SNA literature, as it  contains some deviations from the structure of the network, identified with the CPM Main path method.  Figure~\ref{keyRoute} shows the obtained Key-route paths, which contain 127 nodes. Basically, we still get the division into three previously mentioned groups. \medskip   

Key-route paths \textbf{the first period (1944--1999)} includes 50 works of the SNA from the social scientists. It starts with two works of Heider on his \textit{theory of social perception and cognitive organization} (1944, 1946), which form the basis for the work of Cartwright (1956) on \textit{structural balance}. Later, two works of Holland on \textit{structural models} follow, published in 1970--1971. Next comes the classical paper of Granovetter on \textit{strength of weak ties} (1973), which is a basis for the works of Breiger on \textit{clustering relational data} and White on \textit{blockmodels}, followed by Alba on the \textit{measure based on social proximity} in networks, and Boorman on \textit{role structures in multiple networks}, published in 1975--76. Then there are 6 works of Burt on \textit{positions in multiple networks (stratification and prestige), structural equivalence and networks subgoups}, published from 1977 to 1981, which have connections to  the works of Holland on \textit{social structure}, Breiger, Lauman, and Wellman on {communities structures}, Breiger on \textit{social roles}, and Faust on \textit{structural and general equivalences}, published at about the same time period. Summing up, this group of works is dealing with \textbf{network and community structures, positions, structural equivalence, and blockmodels}.  \medskip 

These works are followed by works on \textbf{measurement and different network metrics}: of Romney and Bernard (1982) on \textit{recalled data for networks construction}, and Stephenson on \textit{centrality} (1989). The last work is also connected to the works of Mizruchi on \textit{measures of influence}, Bonacich on \textit{power and centrality measures}, and Burt, Mariolis, Mizruchi on \textit{interlock networks}. This is followed by the work of Freeman on the \textit{measures of centrality}, which was published in 1991, and it is very strongly connected to the work of Valente on \textit{social network thresholds in the diffusion of innovations} (1996). Another strong connection of Valente goes to the previous work of Michaelson (1993) on the \textit{development of a scientific specialty as a diffusion through social relations}.  \medskip 

The work of Valente is the one bridging the first group of scientists from the social science with the \textbf{group of physicists}, which includes 28 works from the Network science discipline published in the \textbf{second period (1999--2008)}. Valente`s work was cited by Newman in the work on the \textit{small-world network model}, appeared in 1999. This work is followed by others on the same topic (small-world networks), written by Moore, Newman, as well as by the work of Callaway on \textit{random graphs} (2000). Then both directions meet at the work of Strogatz on \textit{complex networks}. Then this topic continues, including \textit{clustering and preferential attachment in growing networks and spread of epidemic diseases on networks} (Newman, 2001, 2002). Since 2003 to 2006, this topic goes to the direction of \textit{community structures detection in large networks}. \medskip 
 
We should note, however, that there is also an \textbf{epidemiological turn} in the observed network, which starts from the works of Stephens and Freeman, followed by Milardo, Neaigus, and Rothenberg in the works on the \textit{diseases transmission} (1992--98), and Potterat in the \textit{infections transmission} (1999). These works are cited by Ferguson (disease transmission), and then the route comes back to the main path to the Newman`s work on the structure and function of complex networks (2003). \medskip 

Since that time, the topics of the obtained Key-routes network change significantly. The work of Newman on community structures is strongly connected to the work of Lusseau (2009) on \textbf{animal social networks}, which starts the \textbf{third period (2008--2018)} with 49 works of the behavioural ecologists. This work is followed by many others, on the same topic: Krause, James (2009) with \textit{general works} on animal SNA, and Ramos-Fernandez, Kasper, Voell, Lehmann, Brent, Sueur (2009--2011), working with \textit{social networks of Nonhuman Primates} (monkeys, baboons). These works are followed by Croft (2011), which represent a practical guide on \textit{hypothesis testing} in Animal social networks. This work is cited by others presenting the research on \textit{mixed-species groups} (Farine), \textit{killer whales} (Foster), \textit{sharks} (Mourier), \textit{dolphins} (Cantor), published in 2012, and \textit{birds} (Silk) and \textit{starlings} (Boogert), published in 2014. There are also some more works on the \textit{methodological issues} -- of Hobson (\textit{An analytical framework for quantifying and testing patterns of temporal dynamics in social networks}), Castels (\textit{Social networks created with different techniques are not comparable}), and Pinter-Wollman (\textit{The dynamics of animal social networks: analytical, conceptual, and theoretical advances}), published in 2013-2014. These works are followed by four works of Farine, published in 2015, on both \textit{methodological issues on constructing, conducting and interpreting animal SNA}, and study of the \textit{wild birds territory acquisition}. We should also note that there are some works connected to the main path, which represents the \textit{social personality and phenotype} (Wilson, Alpin, Farine), published in 2013-2014.\medskip   
 
The upper part of the network contains of works published in the last years, 2016--2018. It presents studies on \textit{disease transmission} (Adelman, Sah, Silk, Dougherty), and the studies of \textit{animal paths tracking} (Leu, Spiegel). Also it contains works on \textit{theoretical issues} (\textit{Current directions in animal social networks} by Croft, \textit{Social traits, social networks and evolutionary biology} by Fisher) and \textit{implementation of different models of network analysis to Animal behaviour research}:  exponential random graph models and statistical network models (Silk), the potential of stochastic actor-oriented models (Fisher), dynamic vs. static SNA (Farine). \medskip   

The full information on the papers (first author, title, journal, year of publication) included into the Main path and Key-route paths is presented in Table~\ref{compareA} in the Appendix C. It is also relevant for our analysis of the islands, presented in the following subsections. In this table, the second column (code) describes in which analysis the work appears: 1- Key-routes, 2- Main Path (CPM), 3- Island 5, 4 - Island 4, 5 - Node Island, 6 - Probabilistic Flow Island. 
 
\subsection{Link Islands}

\begin{figure}
\begin{center}
\includegraphics[width=\textwidth,viewport=45 0 615 695,clip=]{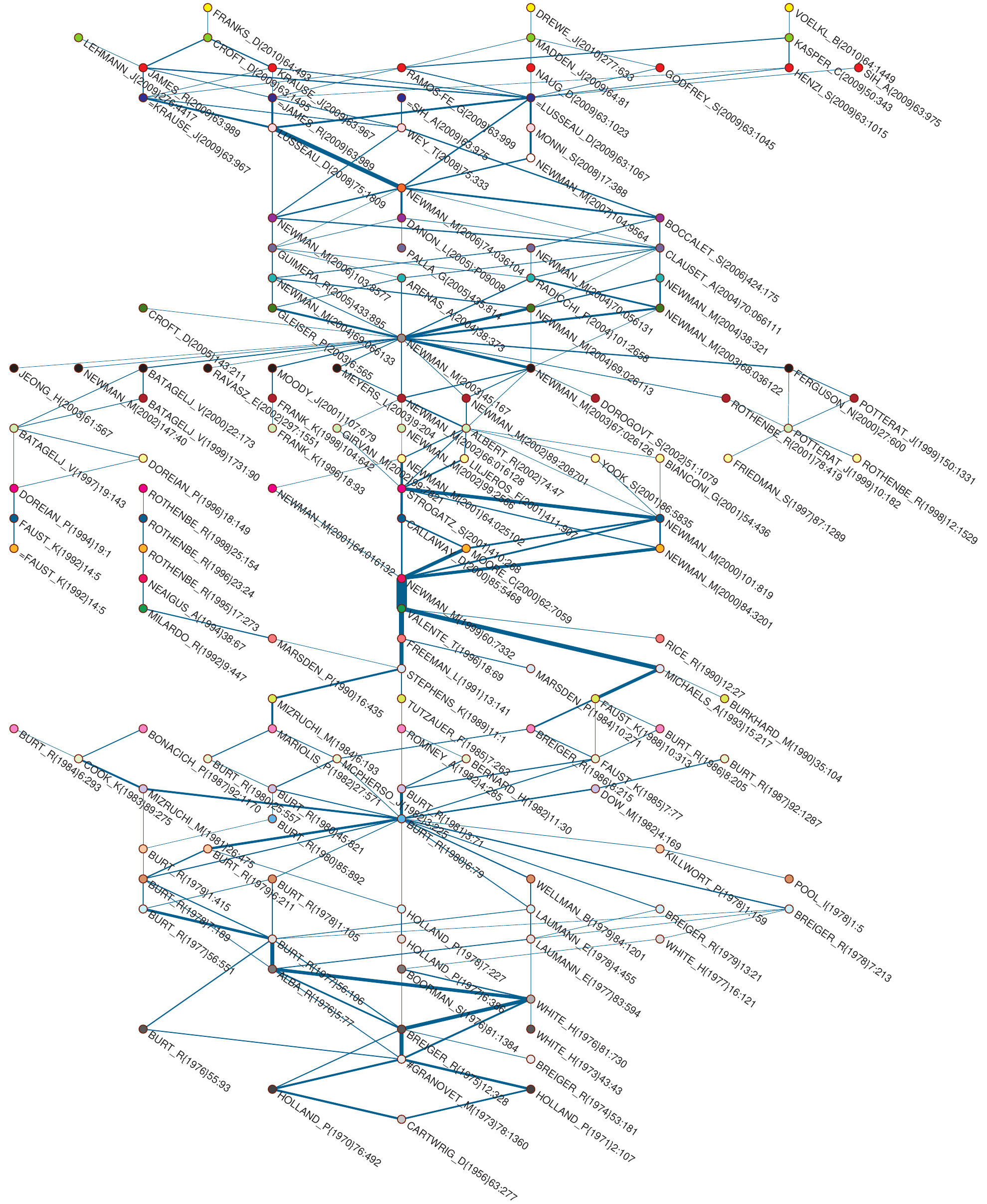}
\end{center}
\caption{SPC net: Island 4} \label{island4}
\end{figure}

Using Islands approach, we searched for SPC link islands (on link weights) \cite[p.~55--57]{Understand} with the number of nodes between 10 and 200, and found 5 islands of 138, 65, 13, 12, and 11 nodes. The obtained largest Island 4 with 138 nodes is presented in Figure~\ref{island4}. Its structure reminds the structure of the Key-route paths - there are 89 overlapping nodes in two networks. The majority of the works presented in this island (from bottom to the work of Valente, published in 1996) belong to the social network scientists, whose works were alreday discussed in previous subsection. In comparison to the Key-routes, this network includes more evident group of \textit{works on blockmodeling} -- by Faust, Doreian, and Batagelj, published in 1992--1997. In the physicists part (from Newman, 1999 to Newman, 2006 on the main route) the topic of \textit{evolving networks} is also presented (Bianconi, Yook, 2001, Jeong, 2003). The third, behavioural ecologists` part is pretty short and finishes by the works on animal social networks published in 2010.  \medskip   

However, this group is fully presented in another Island 5 containing 65 nodes (Figure~\ref{island4}). It has 39 overlapping nodes with the Key-routes. `New' works in the island also belong to the topics on animal social networks described above. However, there are some works devoted to the methodological issues of Network analysis itself --  \textit{reconstructing animal social networks from independent small-group observations} (Perreault, 2010), \textit{temporal dynamics and network analysis} (Blonder, 2012), \textit{mining of animal social systems} (Krause, 2013), \textit{animal social network inference and permutations for ecologists in R} (Farine, 2013), \textit{estimating uncertainty and reliability of social network data using Bayesian inference} (Farine, 2015). It is interesting that this group form a separate subnetwork, even though it is connected to the upper part of Island 4 by topic. It may mean that the works included into this subnetwork are more connected to each other, while social animal network works in Island 4 are more strongly connected to the previous works of physicists.  \medskip

\begin{figure}
\begin{center}
\includegraphics[width=\textwidth,viewport=25 55 595 330,clip=]{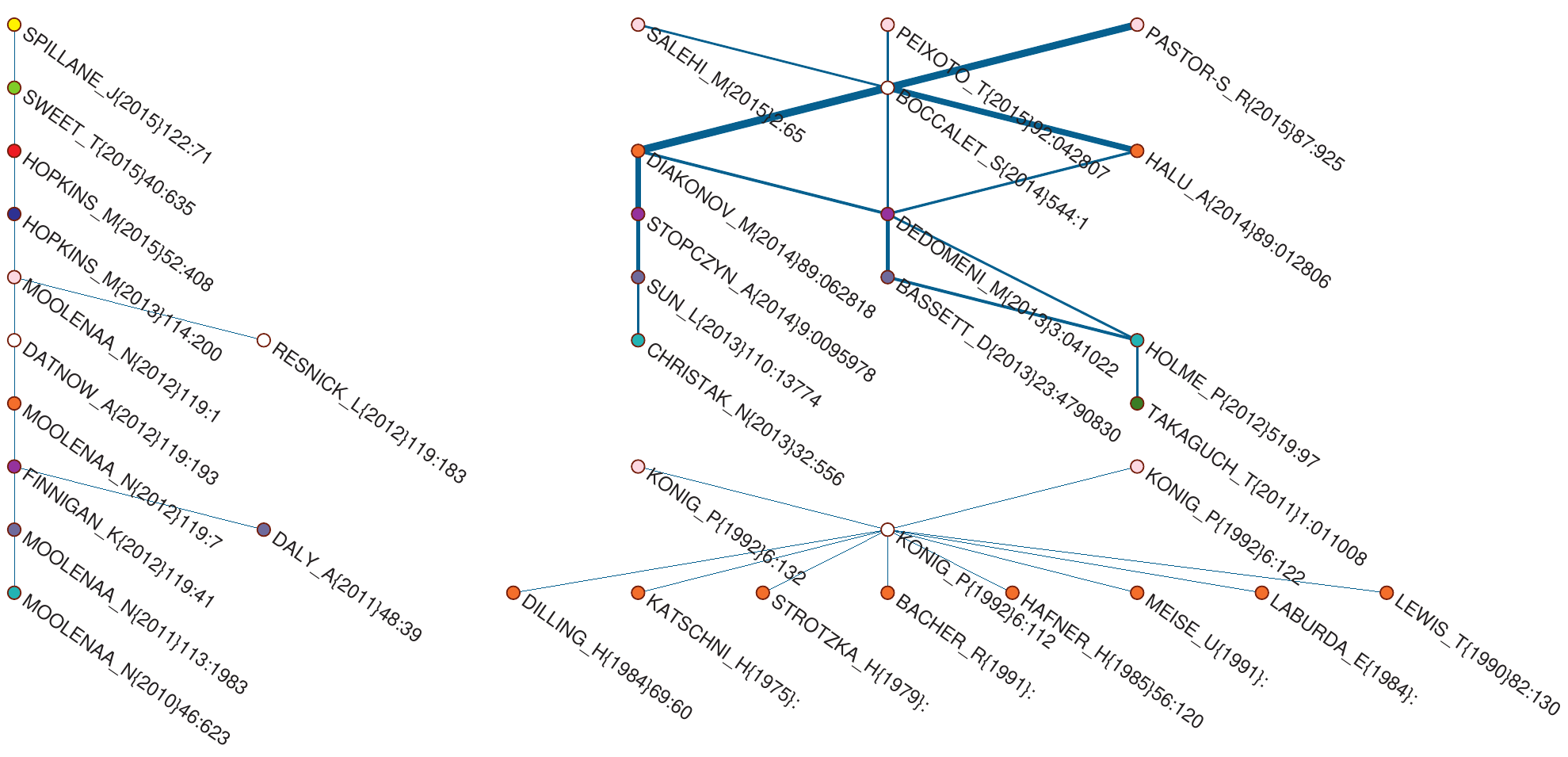}
\end{center}
\caption{SPC net: Islands 1-3} \label{citeisl1-3}
\end{figure}

Other obtained islands are presented in Figure~\ref{citeisl1-3}. For the purpose of better visibility of the picture, the weights were multiplied by 100. The left Island 2 consists of 12 works in the field of social networks in \textit{education}, including issues of leadership, teachers and students communication and collaboration. Another very coherent group is presented in the same figure on the bottom left. These are 11 works in \textit{Neuropsychiatrie} written by Austrian authors. The left upper island presents 13 works of \textit{physicists} with the strongest links between the work of Boccaletti published in 2014 on the structure and dynamics of multilayer networks and others on the topics of complex, multilayer, dynamic, and temporal networks, as well as spreading processes in these networks.  \medskip    

\begin{figure}
\begin{center}
\includegraphics[width=\textwidth,viewport=150 5 585 420,clip=]{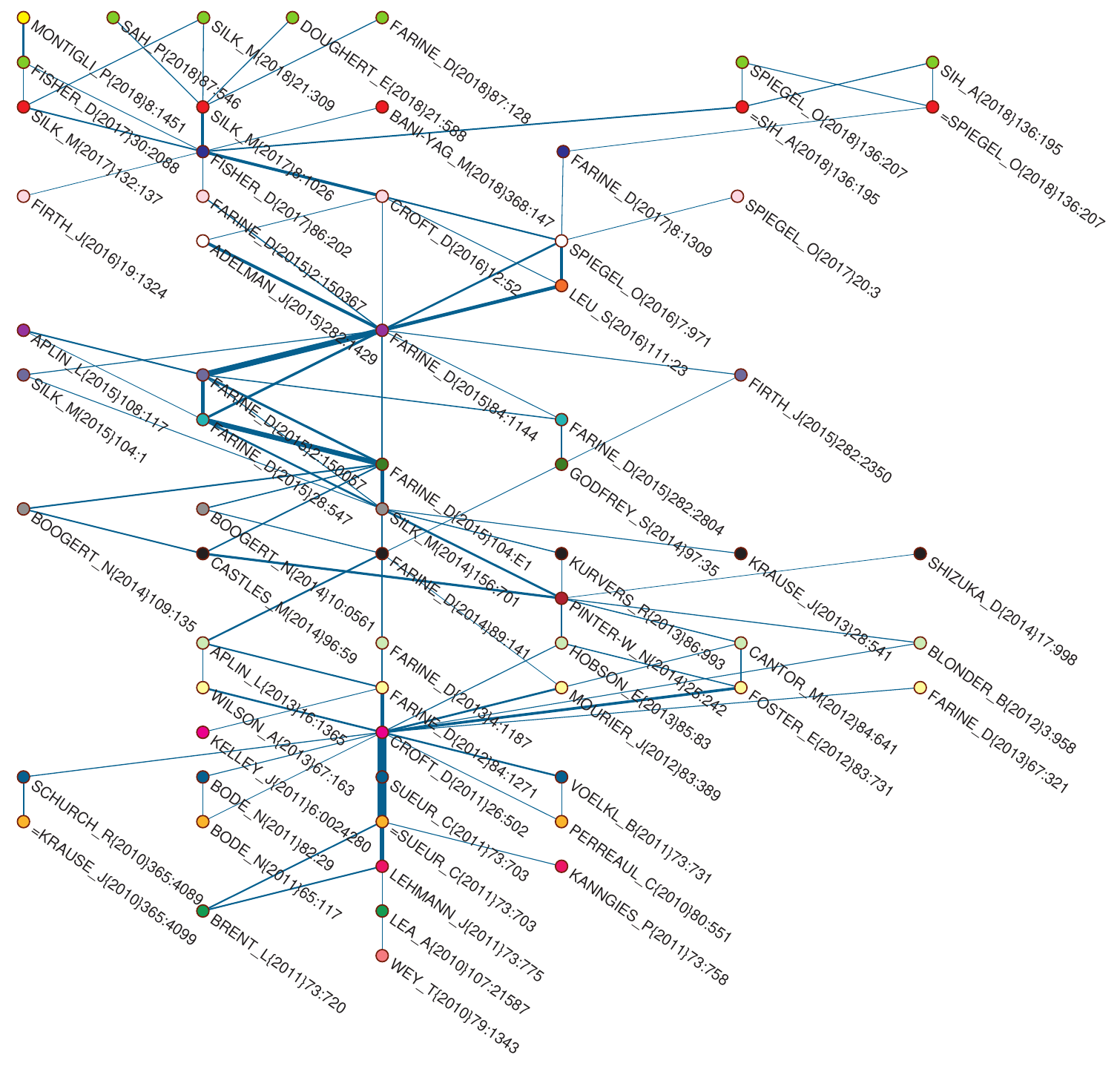}
\end{center}
\caption{SPC net: Island 5} \label{island5}
\end{figure}

Using \textbf{Node islands approach} we searched for the node islands in SPC network of size [10, 200], and got one island of size 200. The works appeared in that island in large part overlap with the works from Islands 4 and 5. These works are listed with the code 5 in the Table~\ref{compareA}.


\subsection{Probabilistic flow}

The Probabilistic flow algorithm determines a node index (and a link weight). The node value is equal to the probability that a random path starting in some initial node reaches this node. We computed the Probabilistic flow \cite[p.~81--82]{Understand} on the $\mathbf{CiteT}$ network, and determined the 200 nodes with the largest values of the probabilistic flow index. They are presented in Table~\ref{compareA}.

39 works out of the first 60 works from the obtained table overlap with the works from the table of 60 works with the largest indegree values of \textbf{CiteN} network (Table~\ref{mostcited}). Half of the listed works with the largest weights in Probabilistic flow network (except \texttt{O`REILLY\_T(2005)} on web 2.0, and \texttt{ALBERT\_R(1999)401:130} on world-wide web) are presented in the most cited works list. The largest values of probabilistic flow are already presented in the indegree distribution of \textbf{CiteN} network: works of \textit{Wasserman and Faust (1994), Watts (1998), Granovetter (1973), Boyd (2007), Barabasi (1991), Freeman (1979), Burt (1992), Milgram (1967)}. There are also physicists in the top of this distribution are \textit{Watts (1998), Barabasi (1999), Girvan (2002), Newman (2003), Albert (2002)}. Works appeared in the list of probabilistic flow, which are not in the list of the most cited works, are works of physicists (Strogatz, Watts, Albert), computer scientists (Brin), mathematicians (Bollobas), scientometricians (Page, Redner), and social scientists (Katz, Mitchell, Glaser).  

By contrast, the obtained set of works is quite different from the lists of most ``important'' works obtained with SPC algorithm (main path and key-routes) and islands approach. However, there are some intersections of the works from the Probabilistic flow list with works which are included to the subnetworks of main path, key-route, and islands 5 and 4 (see Table~\ref{compareA}, Appendix C). The 14 works which appear in the maximum subnetworks, including probabilistic flow, are works of several social scientists -- \textit{Granovetter (1973) on strenght of weak ties, White (1976) on  the blockmodels of roles and positions, and Cartwright (1956) on structural balance and generalization of Heider theory}, -- while the majiority of works belongs to physicists: \textit{Newman, Albert, Strogatz, Clauset, Boccaletti on complex networks and community detection}.  
 
\section{Conclusions}

Our study uses the bibliometric approach for studying the field of SNA. In this paper we presented only the first part of the study --  the analysis of the basic networks constructed out of the collected dataset and their reduced versions, including only the works with the full WoS description. In general, we can make a conclusion on the relevance of the obtained data to the research objects: the lists of most cited works, most used journals and, especially, keywords (with top words \textit{social, network} and \textit{analysis}) do not contradict our basic knowledge of the SNA field. These data were used for more complex analysis. \medskip 

The results show that starting from its institutionalization in the 1980-1990's, SNA field has grown significantly both in terms of the number of publications and the amount of disciplines involved into the research using SNA approach. The number of publications shows the constant growth, and on average it doubles every 3 years.  \medskip 

The analysis confirmed the previous studies on the SNA field development using citation network analysis. Up to the middle of 1990's the most ``important'' works belong to the authors from the \textit{social sciences}, and starting from 2000's the field experience the ``invasion of physicists''. To our surprise, from 2010's both groups experience the ``invasion'' of scientists from a completely another field -- \textit{animal SNA}. The presence of this group is also seen in other results: we identified the journal \textit{Animal Behaviour}, as well as some active authors, having large amount of works. This does not mean that either social scientists or physicists are not presented in the field anymore -- it means that the newly appeared group is quite active both in number of publications and citations of each other. According to the analysis of journals, another active field of SNA research goes from the field of \textit{Computer science}, with \textit{Lecture notes in Computer Science} being the journal with the largest amount of works published. One can argue, however, that this is more a \textit{series} of different publications on Computer Science, including conference proceedings, but not a single journal.\medskip 

However, in spite of all ``invasions'' the most cited works still belong to the social scientists -- with Wasserman, Faust, and Granovetter on the top. Other highly cited works are intermixed between social scientists (Freeman, McPherson, Burt, Coleman, Putnam, Scott, Everett and Borgatti, and others) and physicists (Newman, Watts, Barabasi, Albert, Girvan, and others). \textit{Social networks}, the main journal in the SNA field, occupies a very high position among the journals where the works from our data set were published. It has lower position in terms of citations from the whole data set. \medskip 

Possible explanation of some groups appearance can be due to the nature of algorithms used for identification of main subgroups of the observed citation networks. Main path algorithm \textit{forces} to connect the nodes in the network, even if the line weight between some of them can be low. Islands approach identify locally important part of the network, which should be \textit{distinct} from their neighbourhood. We can propose that the works on some topics could not form a separate island, as they are embedded to the subgroup of main island. More detailed explanation of the different groups in SNA field appearance and coexistence should be provided with the further analysis of derived networks, such as networks of co-authorship and co-citation between journals and authors (Part 2), and temporal analysis of these networks (Part 3).  \medskip 

Once again, we should highlight that for the results of bibliographic network analysis the coverage of bibliographic database used in the research is extremely important. We can propose that for future analysis a combination of different data bases (such as \textit{WoS, Scopus, Google Scholar}, and others) can be used. However, the problem of identification of different entities (journals and authors) can still occur, that is why we can state the need of standardization of information published in bibliographic data bases. \medskip 

\section*{Acknowledgments}
This work is supported in part by the Slovenian Research Agency (research program P1-0294 and research projects J1-9187 and J7-8279) and by Russian Academic Excellence Project '5-100'.

\appendix
\section{Appendix: Synonyms}

Some problems associated with names recognition can occur in the data base. It can happen that the same work is named by different short names. For example, the short names \texttt {BOYD\_D(2007)13} and \texttt {BOYD\_D(2008)13:210} referencing the same work of Danah Boyd, are originally published in 2007, but in many cases referenced as being published in 2008. There were also cases when the short names were different due to the discrepancies in the descriptions -- such as \texttt{GRANOVET\_M(1973)78:1360} and \texttt{GRANOVET\_M(1973)78:6}, or \texttt{COLEMAN\_J(1988)94:95} and \texttt{COLEMAN\_J(1988)94:S95}. Also the names of some authors were presented in a different way -- for example, \texttt{GRANOVET\_M} and \texttt{GRANOVET\_}. We identified these cases for all works with the large (at least 3) indegree frequencies in the Cite network.\medskip 

To resolve these problems, we have to correct the data. There are two possibilities: (1) to make corrections in the local copy of original data (WoS file); and (2) to make an equivalence partition of nodes and shrink the set of works accordingly in all  obtained networks. We used the second option \citep{Understand}. For the works with the large frequences we prepared lists of possible equivalents and manually determined equivalence classes. With a function in R we produced a \textbf{Pajek}'s partition of equivalent work names representing the same work. We used this partition to shrink the networks $\mathbf{Cite, WA, WJ}$, and $\mathbf{WK}$. The partitions $\mathbf{year,  DC}$ and the vector $\mathbf{NP}$ were also shrunk.  \medskip 

Similar problem was also with journals titles. The network \textbf{WJ} had 70,425 journals. Due to the inconsistencies in titles writing in different descriptions, it contained sets of nodes denoting \textit{the same journal}. To get the list of these nodes, we constructed for each journal title a short code, which was formed out of the first two letters of each word in the journal's title, -- such as \texttt{SONEANMI} for \texttt{SOCIAL NETWORK ANALYSIS AND MINING}, -- and then sorted so that the journals with the same code were grouped together. We decided to manually inspect all journals with at least one of their names cited at least 200 times. To get these counters we computed in Pajek the 2-mode network \textbf{Cite*WJc} and determined the vector \textbf{wIndegJ.vec} with weighted indegrees for journals. We obtained a list of candidates for inspection with 5,482 titles. To additinally reduce the number of titles to inspect we decided to consider only titles that appeared in at least 3 citations. Finally, we got the list \textbf{journalK100.csv} with 3,714 titles, that were manually inspected. After manualy checking this list was reduced to 1,688 titles. Some examples of the journal titles grouped according to their codes are presented in the Figure~\ref{jour1}.\medskip 

However, some jurnal titles can appear also in an abbreviated form based on initials -- for example, the \textit{Journal of the American Statistical Association} could be coded as \texttt{JAMSTAS} according to its short title \texttt{J AM  STAT ASS} and as \texttt{JA} according to its abbreviation \texttt{JASA}. That is why we also produced a list of frequent journals names of length at most 5, have chosen all the cases that could be considered as abbreviations, such as \texttt{CACM, JACM, JASA, LNCS, NIPS, JASSS, IJCAI, BMJ, JOSS}, and others, and performed a manual search for the abbreviations of these jornals in the original list of 70,425 journals. We grouped all the jornal titles which included the same abbreviations -- an example  is presented on the Figure~\ref{jour2} (it is seen that there were different codes generated to different titles). The results of the search were added to the first obtained list, and finally the list and the corresponding partition for network shrinking were produced.  \medskip 

\begin{figure}
\renewcommand{\baselinestretch}{0.8}
\scriptsize
\begin{verbatim}
63656		1312696	10849	SONEANMI	| SOCIAL NETWORK ANAL
63657		1330776	3	SONEANMI	| SOCIAL NETWORKS ANAL
63658		1311789	645	SONEANMI	| SOC NETW ANAL MIN
63659		1313366	7	SONEANMI	| SOCIAL NETW ANAL MIN
63660		1315722	7	SONEANMI	| SOC NETW ANAL MINING
...
25340		1297450	195	HUREMA	| HUM RESOURCE MANAGE
25341		1298839	189	HUREMA	| HUMAN RESOURCE MANAG
25343		1304542	3	HUREMA	| HUMAN RESOURCES MANA
25344		1305503	67	HUREMA	| HUM RESOUR MANAGE
25345		1312370	222	HUREMA	| HUM RESOUR MANAGE-US
25352		1301632	189	HUREMAR	| HUM RESOUR MANAGE R
25353		1303129	5	HUREMAR	| HUM RESOUR MANAG R
...
4188		1299141	391	AMJGEPS	| AM J GERIAT PSYCHIAT
4189		1299905	23	AMJGEPS	| AM J GERIATRIC PSYCH
4190		1302259	12	AMJGEPS	| AMER J GERIATR PSYCHIATR
4191		1304932	14	AMJGEPS	| AM J GERIATR PSYCHIA
4192		1314551	7	AMJGEPS	| AM J GERIATR PSYCHIATRY
\end{verbatim}
\caption{An example of different journals titles writing}\label{jour1}
\end{figure}

\begin{figure}
\renewcommand{\baselinestretch}{0.8}
\scriptsize
\begin{verbatim}
10524		1297183	50		5912	COAC		| COMMUN ACM
10525		1311274	14141		6	COAC		| COMMUNICATIONS ACM
10062		1309889	12756		61	CA		| CACM
...
55366		1351847	54714		1	PSPOSC 	| PS POLITICAL SCIENCE
55768		1320199	23066		5	POSC		| POLITICAL SCI
55769		1320573	23440		3	POSC		| POLIT SCI
56082		1297982	849		224	PSSCPO	| PS-POLIT SCI POLIT
56083		1298064	931		110	PSSCPO	| PS-POLITICAL SCI POL
...
33087		1299216	2083		1617	JAC		| J ACM
33550		1355703	58570		2	JACJA		| J ACM JACM
32955		1302464	5331		17	JA		| JACM
\end{verbatim}
\caption{An example of different titles journals writing with abbreviations}\label{jour2}
\end{figure}

\section{Appendix: Strong components}

\begin{figure}
\begin{center}
\includegraphics[width=\textwidth,viewport=75 34 690 535,clip=]{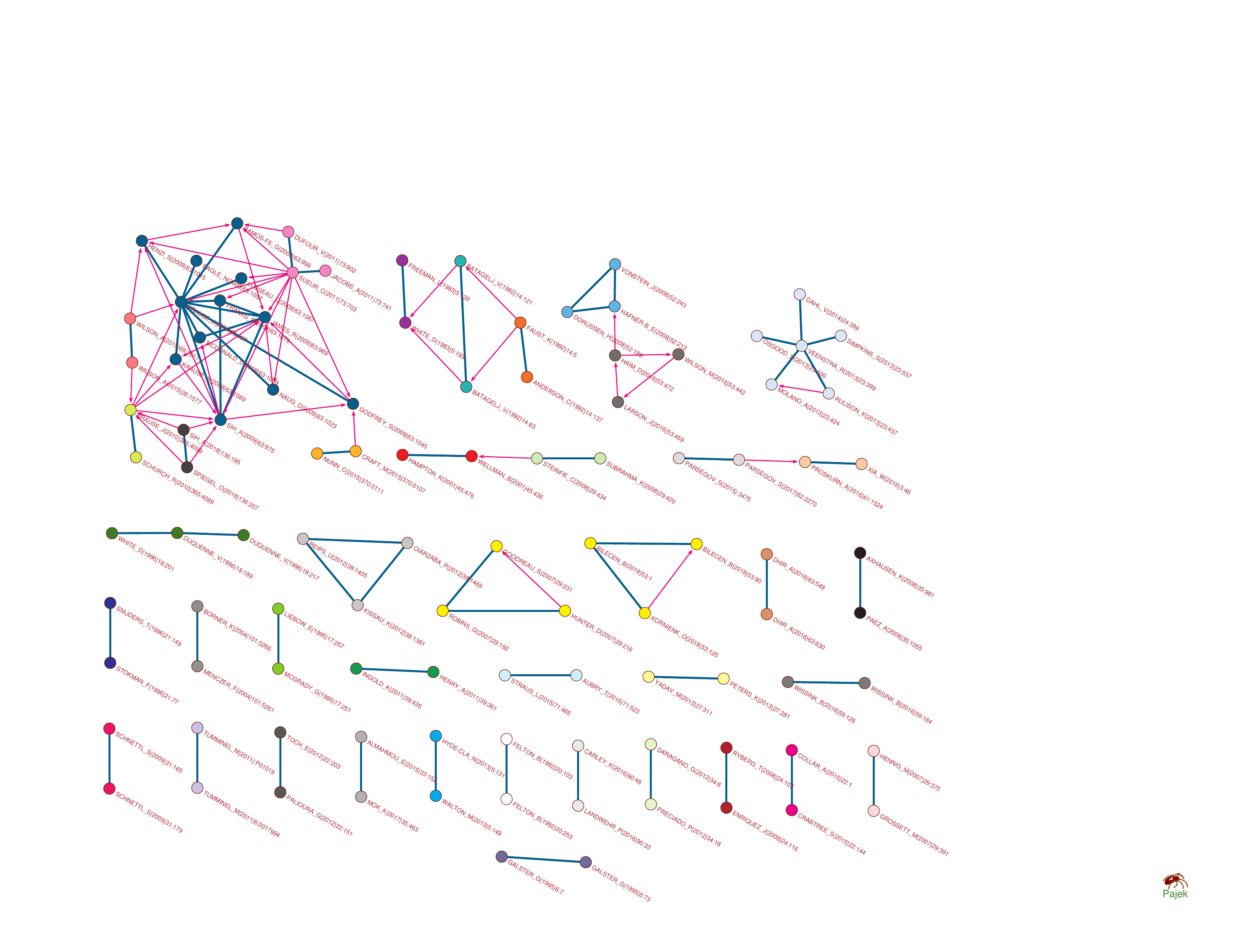}
\end{center}
\caption{SPC net: Strong components} \label{citecomp}
\end{figure}

The citation network \textbf{CiteB} has 41 nontrivial strong components of different sizes, which are presented in the  Figure~\ref{citecomp}. The reciprocal (cycle) links are marked with the bluse colour, while directed pink lines also show the connections of these nodes with others. In the majority of cases, mutual referencing between the works is a characteristic of papers published in the same issue of the journal. For example, the first large cycle is combined of 12 works published in a special issue named \textit{Social Networks: new perspectives} in the journal \textit{Behavioral Ecology and Sociobiology} (Volume 63, Issue 7, May 2009). Another example are the works \texttt {BATAGELJ\_V(1992)14:63} and \texttt {BATAGELJ\_V(1992)14:121}, and \texttt {FAUST\_K(1992)14:5} and \texttt {ANDERSON\_C(1992)14:137} in the special issue on \textit{Blockmodels} in the journal \textit{Social networks} (Volume 14, Issues 1–2, March–June 1992). \medskip

Other cases are connections due to the same author (\texttt {TUMMINEL\_M(2011):P01019} and \\
 \texttt {TUMMINEL\_M(2011)6:0017994}, \texttt{WILSON\_A(2015)69:1617} and \texttt{WILSON\_A(2015)26:1577}, \texttt { PARSEGOV\_S(2015):3475} and \texttt {PARSEGOV\_S(2017)62:2270}) or journal ( \texttt {VEENSTRA\_R(2013)23:399} and \texttt {DAHL\_V(2014)24:399}). However, there are cases when the authors and journals of publications are different (\texttt {ALMAHMOU\_E(2015)33:152} and \texttt {MOK\_K(2017)35:463}, \texttt {XIA\_W(2016)3:46} and \texttt {PROSKURN\_A(2016)61:1524}). \medskip

\section{Appendix: Main publications}
\begin{landscape}
\LTcapwidth=220mm
\denseFont
\begin{longtable}{p{0.7cm}|p{0.8cm}|p{3cm}|p{14.5cm}|p{3.5cm}l}
\caption{Citation CiteT net: \label{compareA} Overlapping of components: (1-Key Routes, 2-CPM Main Path, 3-Island 5, 4-Island 4, 5-Node Island, 6-Probabilistic Flow)} \\
\renewcommand{\arraystretch}{0.7}
year& 	 code& 	 author& 	 title& 	jour or book\\ \hline \endhead
1934& 	6& 	 Moreno JL& 	 Who Shall Survive: A New Approach to the Problem of Human Interrelations& 	book\\
1941& 	6& 	 Davis A & 	 Deep South: A Social Anthropological Study of Caste and Class& 	book\\
1944& 	12& 	 Heider F& 	 Social perception and phenomenal causality& 	 psychol rev\\
1946& 	12& 	 Heider F& 	 Attitudes and cognitive organization& 	 j psychol\\
1948& 	6& 	 Bavelas A& 	 A mathematical model for group structure& 	 hum organ\\
1950& 	6& 	 Homans GC& 	 The human group& 	 book\\
1951& 	6& 	 Leavitt HJ& 	 Some effects of certain communication patterns on group performance& 	 j abnorm soc psych\\
1953& 	6& 	 Katz L& 	 A new status index derived from sociometric analysis& 	 psychometrika\\
1954& 	6& 	 Barnes JA& 	 Class and committees in a norwegian island parish& 	 hum relat\\
1955& 	6& 	 Katz E& 	 Personal influence& 	 book\\
1956& 	12456& 	 Cartwright D& 	 Structural balance - a generalization of Heider theory& 	 psychol rev\\
1957& 	6& 	 Bott E& 	 Family and social network: roles& 	 book\\
1958& 	6& 	 Heider F& 	 The psychology of interpersonal relations& 	 book\\
1959& 	6& 	 Goffman E& 	 The presentation of self in everyday life& 	 book\\
1959& 	6& 	 Erdos P& 	 On random graphs I& 	 book\\
1960& 	6& 	 Erdos P& 	 On the evolution of random graphs& 	 publ mat inst has\\
1962& 	6& 	 Rogers EM& 	 Diffusion of innovations& 	 book\\
1965& 	6& 	 Price DJD& 	 Networks of scientific papers& 	 science\\
1965& 	6& 	 Harary F& 	 Structural models: an introduction to the theory of directed graphs& 	 book\\
1965& 	6& 	 Hubbell CH & 	An input-output approach to clique identification& 	 sociometry\\
1966& 	6& 	 Sabidussi G& 	 The centrality of a graph& 	 book\\
1966& 	6& 	 Coleman JS& 	 Equality of educational opportunity& 	 book\\
1967& 	6& 	 Glaser BG& 	 The discovery of grounded theory: strategies for qualitative theory& 	 book\\
1967& 	6& 	 Milgram S& 	 The small world problem& 	 psychol today\\
1967& 	6& 	 Milgram S& 	 The small world problem& 	 book\\
1969& 	6& 	 Travers J& 	 An experimental study of the small world problem& 	 book\\
1969& 	6& 	 Kauffman S& 	 Metabolic stability and epigenesis in randomly constructed genetic nets& 	 theoret biol \\
1969& 	6& 	 Mitchell JC& 	 Social networks in urban situations: analyses of personal relationships in central african towns& 	 book\\
1970& 	1245& 	 Holland PW& 	 Method for detecting structure in sociometric data& 	 amer j sociol\\
1970& 	5& 	 White HC& 	 Search parameters for small world problem& 	 soc forces\\
1970& 	6& 	 Kernighan BW& 	 An efficient heuristic procedure for partitioning graphs& 	 book\\
1971& 	145& 	 Holland PW& 	 Transitivity in structural models of small groups& 	 comp group stud\\
1971& 	6& 	 Lorrain F & 	 Structural equivalence of individuals in social networks& 	 book\\
1972& 	6& 	 Bonacich P& 	 Factoring and weighting approaches to status scores and clique identification& 	 j math sociol\\
1973& 	12456& 	 Granovetter MS& 	 Strength of weak ties& 	 amer j sociol\\
1973& 	4& 	 White HC& 	 Everyday life in stochastic networks& 	 sociol inq\\
1973& 	5& 	 Holland PW& 	 Structural implications of measurement error in sociometry& 	 j math sociol\\
1973& 	6& 	 Laumann EO& 	 Bonds of pluralism: the form and substance of urban social networks& 	 book\\
1974& 	45& 	 Breiger RL& 	 Duality of persons and groups& 	 soc forces\\
1974& 	6& 	 Granovetter MS& 	 Getting a job: a study of contacts and careers& 	 book\\
1975& 	1245& 	 Breiger RL& 	 Algorithm for clustering relational data with applications to SNA and comparison with multidimensional-scaling& 	 j math psychol\\
1975& 	6& 	 Fishbein M& 	 Intention and behavior: an introduction to theory and research& 	 book\\
1976& 	12456& 	 White HC& 	 Social-structure from multiple networks 1 Blockmodels of roles and positions& 	 amer j sociol\\
1976& 	1245& 	 Alba RD& 	 Intersection of social circles - new measure of social proximity in networks& 	 sociol method res\\
1976& 	145& 	 Burt RS& 	 Positions in networks& 	 soc forces\\
1976& 	145& 	 Boorman SA& 	 Social-structure from multiple networks 2 Role structures& 	 amer j sociol\\
1977& 	1245& 	 Burt RS& 	 Positions in multiple network systems 1 General conception of stratification and prestige in a system of actors cast as a social topology& 	 soc forces\\
1977& 	1245& 	 Burt RS& 	 Positions in multiple network systems 2 Stratification and prestige among elite decision-makers in community of altneustadt& 	 soc forces\\
1977& 	145& 	 Holland PW& 	 Social-structure as a network process& 	 z soz\\
1977& 	45& 	 Laumann EO& 	 Community-elite influence structures - extension of a network approach& 	 amer j sociol\\
1977& 	45& 	 White HC& 	 Probabilities of homomorphic mappings from multiple graphs& 	 j math psychol\\
1977& 	6& 	 Freeman LC& 	 Set of measures of centrality based on betweenness& 	 sociometry\\
1977& 	6& 	 Zachary WW& 	 An information flow model for conflict and fission in small groups& 	 book\\
1978& 	1245& 	 Burt RS& 	 Cohesion versus structural equivalence as a basis for network subgroups& 	 sociol method res\\
1978& 	145& 	 Holland PW& 	 Omnibus test for social-structure using triads& 	 sociol method res\\
1978& 	145& 	 Laumann EO& 	 Community structure as interorganizational linkages& 	 annu rev sociol\\
1978& 	145& 	 Breiger RL& 	 Joint role structure of 2 communities elites& 	 sociol method res\\
1978& 	456& 	 Pool ID& 	 Contacts and influence& 	 soc networks\\
1978& 	45& 	 Killworth PD& 	 Reversal small-world experiment& 	 soc networks\\
1978& 	45& 	 Burt RS& 	 Stratification and prestige among elite experts in methodological and mathematical sociology circa 1975& 	 soc networks\\
1978& 	6& 	 Granovetter M& 	 Threshold models of collective behavior& 	 am j sociol\\
1979& 	1245& 	 Burt RS& 	 Relational equilibrium in a social topology& 	 j math sociol\\
1979& 	145& 	 Wellman B& 	 Community question - intimate networks of east yorkers& 	 amer j sociol\\
1979& 	45& 	 Breiger RL& 	 Toward an operational theory of community elite structures& 	 qual quant\\
1979& 	45& 	 Burt RS& 	 Structural theory of interlocking corporate directorates& 	 soc networks\\
1979& 	6& 	 Freeman LC& 	 Centrality in social networks conceptual clarification& 	 soc networks\\
1979& 	6& 	 Berkman LF& 	 Social networks, host-resistance, and mortality - 9-year follow-up-study of alameda county residents& 	 amer j epidemiol\\
1979& 	6& 	 Garey MR& 	 Computers and intractability: a guide to the theory of NP-completeness& 	 book\\
1980& 	1245& 	 Burt RS& 	 Models of network structure& 	 annu rev sociol\\
1980& 	1245& 	 Burt RS& 	 Testing a structural theory of corporate cooptation - interorganizational directorate ties as a strategy for avoiding market constraints on profits& 	 amer sociol rev\\
1980& 	45& 	 Burt RS& 	 Cooptive corporate actor networks - a reconsideration of interlocking directorates involving American manufacturing& 	 admin sci quart\\
1980& 	45& 	 Burt RS& 	 Autonomy in a social topology& 	 amer j sociol\\
1981& 	145& 	 Mizruchi MS& 	 Influence in corporate networks - an examination of 4 measures& 	 admin sci quart\\
1981& 	145& 	 Burt RS& 	 A note on inferences regarding network subgroups& 	 soc networks\\
1981& 	6& 	 Holland PW& 	 An exponential family of probability-distributions for directed-graphs& 	 j amer statist assn\\
1981& 	6& 	 Feld SL& 	 The focused organization of social ties& 	 am j sociol\\
1982& 	1245& 	 Mcpherson JM& 	 Hypernetwork sampling - duality and differentiation among voluntary organizations& 	 soc networks\\
1982& 	1245& 	 Mariolis P& 	 Centrality in corporate interlock networks - reliability and stability& 	 admin sci quart\\
1982& 	145& 	 Bernard HR& 	 Informant accuracy in social-network data 5 An experimental attempt to predict actual communication from recall data& 	 soc sci res\\
1982& 	145& 	 Romney AK& 	 Predicting the structure of a communications network from recalled data& 	 soc networks\\
1982& 	145& 	 Dow MM& 	 Network auto-correlation - a simulation study of a foundational problem in regression and survey-research& 	 soc networks\\
1982& 	6& 	 Fischer CS& 	 To dwell among friends: personal networks in town and city& 	 book\\
1982& 	6& 	 Burt RS & 	 Toward a structural theory of action: network models of social structure, perception and action& 	 book\\
1983& 	145& 	 Cook KS& 	 The distribution of power in exchange networks - theory and experimental results& 	 am j sociol\\
1983& 	6& 	 Granovetter M & 	The strength of weak ties: a network theory revisited& 	 sociol theory\\
1983& 	6& 	 Salton G& 	 introduction to modern information retrieval& 	 book\\
1984& 	1245& 	 Mizruchi MS& 	 Interlock groups, cliques, or interest-groups - comment& 	 soc networks\\
1984& 	45& 	 Burt RS& 	 Network items and the general social survey& 	 soc networks\\
1984& 	45& 	 Marsden PV& 	 Mathematical ideas in social structural-analysis& 	 j math sociol\\
1984& 	6& 	 Lazarus R& 	 Stress, appraisal, and coping& 	 book\\
1984& 	6& 	 Axelrod R& 	 The evolution of cooperation& 	 book\\
1984& 	6& 	 Kuramoto Y & 	 Chemical oscillations, waves, and turbulence& 	 book\\
1985& 	145& 	 Faust K& 	 Does structure find structure - a critique of burt use of distance as a measure of structural equivalence& 	 soc networks\\
1985& 	145& 	 Tutzauer F& 	 Toward a theory of disintegration in communication-networks& 	 soc networks\\
1985& 	6& 	 Cohen S& 	 Stress, social support, and the buffering hypothesis& 	 psychol bull\\
1985& 	6& 	 Granovetter M& 	 Economic-action and social-structure - the problem of embeddedness& 	 amer j sociol\\
1985& 	6& 	 Bollobas B& 	 Random graphs& 	 book\\
1986& 	145& 	 Breiger RL& 	 Cumulated social roles - the duality of persons and their algebras& 	 soc networks\\
1986& 	45& 	 Burt RS& 	 A cautionary note& 	 soc networks\\
1986& 	6& 	 Bourdieu P & 	 The forms of capital& 	 book\\
1986& 	6& 	 Baron RM& 	 The moderator mediator variable distinction in social psychological-research - conceptual, strategic, and statistical considerations& 	 j personal soc psychol\\
1986& 	6& 	 Bandura A& 	 Social foundations of thought and action: a social cognitive theory& 	 book\\
1987& 	1456& 	 Bonacich P& 	 Power and centrality - a family of measures& 	 amer j sociol\\
1987& 	145& 	 Burt RS& 	 Social contagion and innovation - cohesion versus structural equivalence& 	 amer j sociol\\
1988& 	145& 	 Faust K& 	 Comparison of methods for positional analysis - structural and general equivalences& 	 soc networks\\
1988& 	6& 	 House JS& 	 Social relationships and health& 	 science\\
1988& 	6& 	 Coleman JS& 	 Social capital in the creation of human capital& 	 am jour soc\\
1988& 	6& 	 Wellman B& 	 Social structures: a network approach& 	 *book\\
1989& 	1245& 	 Stephenson K& 	 Rethinking centrality - methods and examples& 	 soc networks\\
1989& 	6& 	 Kamada T& 	 An algorithm for drawing general undirected graphs& 	 inform process lett\\
1989& 	6& 	 Davis FD& 	 Perceived usefulness, perceived ease of use, and user acceptance of information technology& 	 mis quart\\
1989& 	6& 	 Kochen M & 	 The small world& 	 book\\
1990& 	1456& 	 Marsden PV& 	 Network data and measurement& 	 annu rev sociol\\
1990& 	4& 	 Burkhardt ME& 	 Changing patterns or patterns of change - the effects of a change in technology on soc. netw. structure and power& 	 admin sci quart\\
1990& 	4& 	 Rice RE& 	 Individual and network influences on the adoption and perceived outcomes of electronic messaging& 	 soc networks\\
1990& 	6& 	 ColemanJ.& 	 Foundations of social theory& 	 book\\
1990& 	6& 	 Guare J& 	 Six degrees of separation: a play& 	 book\\
1990& 	6& 	 Deerwester S& 	 Indexing by latent semantic analysis& 	 j am soc inf sci tec\\
1991& 	1245& 	 Freeman LC& 	 Centrality in valued graphs - a measure of betweenness based on network flow& 	 soc networks\\
1991& 	6& 	 Ajzen I& 	 The theory of planned behavior& 	 organ behav hum dec\\
1991& 	6& 	 Scott J& 	 Social network analysis: a handbook & 	 book\\
1991& 	6& 	 Lave J & 	 Situated learning: legitimate peripheral participation & 	 book\\
1991& 	6& 	 Fruchterman TMJ& 	 Graph drawing by force-directed placement& 	 software pract exper\\
1992& 	145& 	 Milardo RM& 	 Comparative methods for delineating social networks& 	 j soc person relat\\
1992& 	45& 	 Faust K& 	 Blockmodels - interpretation and evaluation& 	 soc networks\\
1992& 	5& 	 Batagelj V& 	 Direct and indirect methods for structural equivalence& 	 soc networks\\
1992& 	5& 	 Batagelj V& 	 An optimizational approach to regular equivalence& 	 soc networks\\
1992& 	6& 	 Burt RS & 	 Structural holes: the social structure of competition& 	 book\\
1992& 	6& 	 Nowak MA& 	 Evolutionary games and spatial chaos& 	 nature\\
1993& 	145& 	 Michaelson AG& 	 The development of a scientific specialty as diffusion through social-relations - the case of role analysis& 	 soc networks\\
1993& 	6& 	 Putnam RD& 	 Making democracy work: civic institutions in modern italy & 	 book\\
1993& 	6& 	 Padgett JF& 	 Robust action and the rise of the medici, 1400-1434& 	 amer j sociol\\
1993& 	6& 	 Manski CF& 	 Identification of endogenous social effects - the reflection problem& 	 rev econ stud\\
1993& 	6& 	 Ahuja RK& 	 Network flows: theory, algorithms, and applications & 	 book\\
1994& 	145& 	 Neaigus A& 	 The relevance of drug injectors social and risk networks for understanding and preventing hiv-infection& 	 soc sci med\\
1994& 	45& 	 Doreian P& 	 Partitioning networks based on generalized concepts of equivalence& 	 j math sociol\\
1994& 	6& 	 Wasserman S & 	 Social network analysis: methods and applications& 	 book\\
1995& 	145& 	 Rothenberg RB& 	 Choosing a centrality measure - epidemiologic correlates in the colorado-springs study of social networks& 	 soc networks\\
1995& 	6& 	 Molloy M& 	 A critical-point for random graphs with a given degree sequence& 	 random struct algor\\
1995& 	6& 	 Rogers EM& 	 Diffusion of Innovation. 4th& 	 book\\
1995& 	6& 	 Granovetter MS& 	 Getting a Job: A Study of Contacts and Careers& 	 book\\
1995& 	6& 	 Nonaka I& 	 The knowledge creation company: how Japanese companies create the dynamics of innovation& 	 book\\
1995& 	6& 	 Putnam RD& 	 Bowling Alone: America's Declining Social Capital. An Interview with Robert Putnam& 	 j democr\\
1996& 	1245& 	 Valente TW& 	 Social network thresholds in the diffusion of innovations& 	 soc networks\\
1996& 	145& 	 Rothenberg R& 	 The relevance of social network concepts to sexually transmitted disease control& 	 sex transm dis\\
1996& 	45& 	 Doreian P& 	 A partitioning approach to structural balance& 	 soc networks\\
1996& 	4& 	 Frank KA& 	 Mapping interactions within and between cohesive subgroups& 	 soc networks\\
1996& 	6& 	 Wasserman S& 	 Logit models and logistic regressions for social networks 1. An introduction to Markov graphs and p& 	 psychometrika\\
1996& 	6& 	 Kretzschmar M& 	 Measures of concurrency in networks and the spread of infectious disease& 	 math biosci\\
1997& 	45& 	 Friedman SR& 	 Sociometric risk networks and risk for HIV infection& 	 amer j public health\\
1997& 	45& 	 Batagelj V& 	 Notes on blockmodeling& 	 soc networks\\
1997& 	6& 	 Uzzi B& 	 Social structure and competition in interfirm networks: The paradox of embeddedness& 	 admin sci quart\\
1998& 	145& 	 Rothenberg RB& 	 Social network dynamics and HIV transmission& 	 aids\\
1998& 	14& 	 Rothenberg RB& 	 Using social network and ethnographic tools to evaluate syphilis transmission& 	 sex transm dis\\
1998& 	45& 	 Frank KA& 	 Linking action to social structure within a system: Social capital within and between subgroups& 	 amer j sociol\\
1998& 	6& 	 Watts DJ& 	 Collective dynamics of 'small-world' networks& 	 nature\\
1998& 	6& 	 Portes A& 	 Social Capital: Its origins and applications in modern sociology& 	 annu rev sociol\\
1998& 	6& 	 Nahapiet J& 	 Social capital, intellectual capital, and the organizational advantage& 	 acad manage rev\\
1998& 	6& 	 Redner S& 	 How popular is your paper? An empirical study of the citation distribution& 	 book\\
1998& 	6& 	 Wenger E& 	 Communities ofpractice: Learning, meaning, and identity& 	 book\\
1998& 	6& 	 Page L & 	 The pagerank citation ranking: Bringing order to the web.& 	 book\\
1998& 	6& 	 Brin S& 	 The anatomy of a large-scale hypertextual Web search engine& 	 comput networks isdn\\
1998& 	6& 	 Huberman B& 	 Strong regularities in world wide web surfing& 	 science \\
1999& 	1245& 	 Newman MEJ& 	 Scaling and percolation in the small-world network model& 	 phys rev e\\
1999& 	145& 	 Potterat JJ& 	 Chlamydia transmission: Concurrency, reproduction number, and the epidemic trajectory& 	 amer j epidemiol\\
1999& 	145& 	 Potterat JJ& 	 Network structural dynamics acid infectious disease propagation& 	 int j std aids\\
1999& 	45& 	 Batagelj V& 	 Partitioning approach to visualization of large graphs& 	 lect note comput sci\\
1999& 	6& 	 Barabasi AL& 	 Emergence of scaling in random networks& 	 science\\
1999& 	6& 	 Hansen MT& 	 The search-transfer problem: The role of weak ties in sharing knowledge across organization subunits& 	 admin sci quart\\
1999& 	6& 	 Faloutsos M& 	 On power-law relationships of the internet topology& 	 book\\
1999& 	6& 	 Watts DJ& 	 Small Worlds: The Dynamics of Networks Between Order and Randomness& 	 book\\
1999& 	6& 	 Barabasi AL& 	 Mean-field theory for scale-free random networks& 	 physica a\\
1999& 	6& 	 Albert R& 	 Internet - Diameter of the World-Wide Web& 	 nature\\
1999& 	6& 	 Banavar JR& 	 Size and form in efficient transportation networks. Nature,& 	 nature\\
1999& 	6& 	 Kleinberg JM& 	 Authoritative sources in a hyperlinked environment& 	 j acm\\
1999& 	6& 	 Haberman B& 	 Internet: growth dynamics of the world-wide web& 	 nature\\
1999& 	6& 	 Lawrence S& 	 Accessibility of information on the Web. & 	 nature \\
1999& 	6& 	 Barthélémy M & 	 Small-world networks: Evidence for a crossover picture& 	 phys rev lett\\
2000& 	1245& 	 Newman MEJ& 	 Models of the small world& 	 j statist phys\\
2000& 	1245& 	 Moore C& 	 Exact solution of site and bond percolation on small-world networks& 	 phys rev e\\
2000& 	145& 	 Callaway DS& 	 Network robustness and fragility: Percolation on random graphs& 	 phys rev lett\\
2000& 	145& 	 Newman MEJ& 	 Mean-field solution of the small-world network model& 	 phys rev lett\\
2000& 	145& 	 Ferguson NM& 	 More realistic models of sexually transmitted disease transmission dynamics - Sexual partnership networks, pair models, and moment closure& 	 sex transm dis\\
2000& 	45& 	 Batagelj V& 	 Some analyses of Erdos collaboration graph& 	 soc networks\\
2000& 	6& 	 Putnam RD & 	 Bowling alone: America’s declining social capital& 	 book\\
2000& 	6& 	 Jeong H& 	 The large-scale organization of metabolic networks& 	 nature\\
2000& 	6& 	 Berkman LF& 	 From social integration to health: Durkheim in the new millennium& 	 soc sci med\\
2000& 	6& 	 Albert R& 	 Error and attack tolerance of complex networks& 	 nature\\
2000& 	6& 	 Amaral LAN& 	 Classes of small-world networks& 	 proc nat acad sci usa\\
2000& 	6& 	 Broder A& 	 Graph structure in the Web& 	 comput netw\\
2000& 	6& 	 Scott J& 	 Social Network Analysis: A Handbook& 	 book\\
2000& 	6& 	 Shi JB& 	 Normalized cuts and image segmentation& 	 ieee t pattern anal\\
2001& 	12456& 	 Newman MEJ& 	 Clustering and preferential attachment in growing networks& 	 phys rev e\\
2001& 	12456& 	 Strogatz SH& 	 Exploring complex networks& 	 nature\\
2001& 	145& 	 Liljeros F& 	 The web of human sexual contacts& 	 nature\\
2001& 	456& 	 Newman MEJ& 	 Scientific collaboration networks. II. Shortest paths, weighted networks, and centrality& 	 phys rev e\\
2001& 	45& 	 Moody J& 	 Race, school integration, and friendship segregation in America& 	 amer j sociol\\
2001& 	45& 	 Rothenberg R& 	 The risk environment for HIV transmission: Results from the Atlanta and Flagstaff network studies& 	 j urban health\\
2001& 	4& 	 Yook SH& 	 Weighted evolving networks& 	 phys rev lett\\
2001& 	4& 	 Bianconi G& 	 Competition and multiscaling in evolving networks& 	 europhys lett\\
2001& 	6& 	 Mcpherson M& 	 Birds of a feather: Homophily in social networks& 	 annu rev sociol\\
2001& 	6& 	 Newman MEJ& 	 The structure of scientific collaboration networks& 	 proc nat acad sci usa\\
2001& 	6& 	 Lin N& 	 Social capital. A theory of social structure and action.& 	 book\\
2001& 	6& 	 Brandes U& 	 A faster algorithm for betweenness centrality& 	 j math sociol\\
2001& 	6& 	 Domingos P& 	 Mining the network value of customers& 	 book\\
2001& 	6& 	 Goldenberg J& 	 Talk of the network: A complex systems look at the underlying process of word-of-mouth& 	 mark lett\\
2001& 	6& 	 Pastor-Satorras R& 	 Epidemic spreading in scale-free networks& 	 phys rev lett\\
2002& 	12456& 	 Albert R& 	 Statistical mechanics of complex networks& 	 rev mod phys\\
2002& 	12456& 	 Newman MEJ& 	 Spread of epidemic disease on networks& 	 phys rev e\\
2002& 	456& 	 Girvan M& 	 Community structure in social and biological networks& 	 proc nat acad sci usa\\
2002& 	456& 	 Newman MEJ& 	 Assortative mixing in networks& 	 phys rev lett\\
2002& 	45& 	 Dorogovtsev SN& 	 Evolution of networks& 	 adv phys\\
2002& 	45& 	 Newman MEJ& 	 Random graph models of social networks& 	 proc nat acad sci usa\\
2002& 	4& 	 Ravasz E& 	 Hierarchical organization of modularity in metabolic networks& 	 science\\
2002& 	4& 	 Newman MEJ& 	 The structure and function of networks& 	 comput phys commun\\
2002& 	6& 	 Watts DJ& 	 Identity and search in social networks& 	 science\\
2002& 	6& 	 Barabasi AL & 	 Linked: The New Science Of Networks& 	 book\\
2002& 	6& 	 Barabasi AL& 	 Evolution of the social network of scientific collaborations& 	 physica a\\
2002& 	6& 	 Adler PS& 	 Social capital: Prospects for a new concept& 	 acad manage rev\\
2002& 	6& 	 Otte E& 	 Social network analysis: a powerful strategy, also for the information sciences& 	 j inform sci\\
2002& 	6& 	 Richardson M& 	 Mining knowledge-sharing sites for viral marketing& 	 book\\
2003& 	12456& 	 Newman MEJ& 	 The structure and function of complex networks& 	 siam rev\\
2003& 	12456& 	 Newman MEJ& 	 Mixing patterns in networks& 	 phys rev e\\
2003& 	145& 	 Newman MEJ& 	 Why social networks are different from other types of networks& 	 phys rev e\\
2003& 	145& 	 Gleiser PM& 	 Community structure in jazz& 	 adv complex syst\\
2003& 	45& 	 Meyers LA& 	 Applying network theory to epidemics: Control measures for Mycoplasma pneumoniae outbreaks& 	 emerg infect dis\\
2003& 	4& 	 Jeong H& 	 Measuring preferential attachment in evolving networks& 	 europhys lett\\
2003& 	56& 	 Guimera R& 	 Self-similar community structure in a network of human interactions& 	 phys rev e\\
2003& 	6& 	 Rogers EM& 	 Diffusion of innovations& 	 book\\
2003& 	6& 	 Borgatti SP& 	 The network paradigm in organizational research: A review and typology& 	 j manage\\
2003& 	6& 	 Dorogovtsev SN& 	 Evolution of Networks: From Biological Nets to the Internet and WWW& 	 book\\
2003& 	6& 	 Watts DJ& 	 Six Degrees: The Science of a Connected Age& 	 book\\
2003& 	6& 	 Blei DM& 	 Latent Dirichlet allocation& 	 j mach learn res\\
2003& 	6& 	 Adamic LA& 	 Friends and neighbors on the Web& 	 soc networks\\
2003& 	6& 	 Lusseau D& 	 The bottlenose dolphin community of Doubtful Sound features a large proportion of long-lasting associations - Can geographic isolation explain this unique trait?& 	 behav ecol sociobiol\\
2003& 	6& 	 Venkatesh V& 	 User acceptance of information technology: Toward a unified view& 	 mis quart\\
2003& 	6& 	 Kempe D & 	Maximizing the spread of influence through a social network& 	acm sigkdd conf \\
2003& 	6& 	 Kempe D & 	 Maximizing the spread of influence through a social network& 	 acm sigkdd conf \\
2004& 	12456& 	 Newman MEJ& 	 Finding and evaluating community structure in networks& 	 phys rev e\\
2004& 	12456& 	 Newman MEJ& 	 Detecting community structure in networks& 	 eur phys j b\\
2004& 	12456& 	 Clauset A& 	 Finding community structure in very large networks& 	 phys rev e\\
2004& 	1456& 	 Radicchi F& 	 Defining and identifying communities in networks& 	 p natl acad sci usa\\
2004& 	1456& 	 Newman MEJ& 	 Fast algorithm for detecting community structure in networks& 	 phys rev e\\
2004& 	145& 	 Arenas A& 	 Community analysis in social networks& 	 eur phys j b\\
2004& 	145& 	 Newman MEJ& 	 Analysis of weighted networks& 	 phys rev e\\
2004& 	6& 	 Cross RL& 	 The hidden power of social networks: Understanding how work really gets done in organizations& 	 book\\
2004& 	6& 	 Freeman LC& 	 The development of social network analysis. A Study in the Sociology of Science& 	 book\\
2004& 	6& 	 Eubank S& 	 Modelling disease outbreaks in realistic urban social networks& 	 nature\\
2004& 	6& 	 Burt RS& 	 Structural holes and good ideas& 	 amer j sociol\\
2005& 	145& 	 Danon L& 	 Comparing community structure identification& 	 j stat mech-theory e\\
2005& 	456& 	 Guimera R& 	 Functional cartography of complex metabolic networks& 	 nature\\
2005& 	456& 	 Palla G& 	 Uncovering the overlapping community structure of complex networks in nature and society& 	 nature\\
2005& 	4& 	 Croft DP& 	 Assortative interactions and social networks in fish& 	 oecologia\\
2005& 	6& 	 Burt RS& 	 Brokerage and closure: An introduction to social capital& 	 book\\
2005& 	6& 	 Adomavicius G& 	 Toward the next generation of recommender systems: A survey of the state-of-the-art and possible extensions& 	 book\\
2005& 	6& 	 Carrington P& 	 Models and Methods in Social Network Analysis& 	 book\\
2005& 	6& 	 Borgatti SP& 	 Centrality and network flow& 	 soc networks\\
2005& 	6& 	 Gross R& 	 Information revelation and privacy in online social networks& 	 book\\
2006& 	12456& 	 Boccaletti S& 	 Complex networks: Structure and dynamics& 	 phys rep-rev sect phys lett\\
2006& 	12456& 	 Newman MEJ& 	 Finding community structure in networks using the eigenvectors of matrices& 	 phys rev e\\
2006& 	1456& 	 Newman MEJ& 	 Modularity and community structure in networks& 	 proc nat acad sci usa\\
2006& 	6& 	 Kossinets G& 	 Empirical analysis of an evolving social network& 	 science\\
2006& 	6& 	 Newman M & 	 The Structure and Dynamics of Networks& 	 book\\
2006& 	6& 	 Eagle N& 	 Reality mining: sensing complex social systems& 	 pers ubiquit comput\\
2007& 	145& 	 Newman MEJ& 	 Mixture models and exploratory analysis in networks& 	 proc nat acad sci usa\\
2007& 	5& 	 Krause J& 	 Social network theory in the behavioural sciences: potential applications& 	 behav ecol sociobiol\\
2007& 	6& 	 Onnela JP& 	 Structure and tie strengths in mobile communication networks& 	 proc nat acad sci usa\\
2007& 	6& 	 Palla G& 	 Quantifying social group evolution& 	 nature\\
2007& 	6& 	 Christakis NA& 	 The spread of obesity in a large social network over 32 years& 	 n engl j med\\
2007& 	6& 	 Mazer JP& 	 I'll see you on Facebook: The effects of computer-mediated teacher self-disclosure on student motivation, affective learning, and classroom climate& 	 book\\
2007& 	6& 	 Liben-Nowell D& 	 The link-prediction problem for social networks& 	 j am soc inf sci technol\\
2007& 	6& 	 Robins G& 	 An introduction to exponential random graph (p*) models for social networks& 	 soc networks\\
2007& 	6& 	 Fortunato S& 	 Resolution limit in community detection& 	 proc nat acad sci usa\\
2007& 	6& 	 Boyd DM& 	 Social network sites: Definition, history, and scholarship& 	 j comput-mediat comm\\
2007& 	6& 	 Raghavan UN& 	 Near linear time algorithm to detect community structures in large-scale networks& 	 phys rev e\\
2007& 	6& 	 Mislove A& 	 Measurement and Analysis of Online Social Networks& 	 book\\
2007& 	6& 	 Leskovec J& 	 Cost-effective Outbreak Detection in Networks& 	 book\\
2007& 	6& 	 Josang A& 	 A survey of trust and reputation systems for online service provision& 	 decis support syst\\
2007& 	6& 	 Steinfield C& 	The benefits of Facebook friends: Social capital and college students’ use of online social network sites.& 	 j comput-mediat comm\\
2007& 	6& 	 Dwyer C& 	 Trust and privacy concern within social networking sites: A comparison of Facebook and MySpace.& 	 amcis 2007 proc\\
2007& 	6& 	 Lenhart A& 	 Teens, Privacy and  online social networks: how teens manage their online identities and personal information in the age of Myspace& 	book\\
2007& 	6& 	 Ellison NB & 	 The benefits of Facebook “friends:” Social capital and college students’ use of online social network sites& 	 j comput-mediat comm\\
2008& 	1245& 	 Lusseau D& 	 Incorporating uncertainty into the study of animal social networks& 	 anim behav\\
2008& 	145& 	 Wey T& 	 Social network analysis of animal behaviour: a promising tool for the study of sociality& 	 anim behav\\
2008& 	145& 	 Monni S& 	 Vertex clustering in random graphs via reversihle jump Markov chain Monte Carlo& 	 j comput graph stat\\
2008& 	6& 	 Blondel VD& 	 Fast unfolding of communities in large networks& 	 j stat mech-theory e\\
2008& 	6& 	 Smith KP& 	 Social networks and health& 	 annu rev sociol\\
2008& 	6& 	 Gonzalez MC& 	 Understanding individual human mobility patterns& 	 nature\\
2008& 	6& 	 Christakis NA& 	 The collective dynamics of smoking in a large soc.l netw.& 	 new engl j med\\
2008& 	6& 	 Fowler JH& 	 Dynamic spread of happiness in a large soc. netw.: longit. analysis over 20 years in the Framingham Heart Study& 	 brit med j\\
2009& 	1245& 	 Kasper C& 	 A social network analysis of primate groups& 	 primates\\
2009& 	1245& 	 Ramos-FernandezG& 	 Association networks in spider monkeys (Ateles geoffroyi)& 	 behav ecol sociobiol\\
2009& 	1245& 	 Lusseau D& 	 The emergence of unshared consensus decisions in bottlenose dolphins& 	 behav ecol sociobiol\\
2009& 	145& 	 Croft DP& 	 Behavioural trait assortment in a social network: patterns and implications& 	 behav ecol sociobiol\\
2009& 	145& 	 James R& 	 Potential banana skins in animal social network analysis& 	 behav ecol sociobiol\\
2009& 	145& 	 Krause J& 	 Animal social networks: an introduction& 	 behav ecol sociobiol\\
2009& 	145& 	 James R& 	 Potential banana skins in animal social network analysis& 	 behav ecol sociobiol\\
2009& 	145& 	 Krause J& 	 Animal social networks: an introduction& 	 behav ecol sociobiol\\
2009& 	14& 	 Lehmann J& 	 Network cohesion, group size and neocortex size in female-bonded Old World primates& 	 p roy soc b-biol sci\\
2009& 	45& 	 Godfrey SS& 	 Network structure and parasite transmission in a group living lizard, the gidgee skink, Egernia stokesii& 	 behav ecol sociobiol\\
2009& 	45& 	 Sih A& 	 Social network theory: new insights and issues for behavioral ecologists& 	 behav ecol sociobiol\\
2009& 	45& 	 Naug D& 	 Structure and resilience of the social network in an insect colony as a function of colony size& 	 behav ecol sociobiol\\
2009& 	45& 	 Madden JR& 	 The social network structure of a wild meerkat population: 2. Intragroup interactions& 	 behav ecol sociobiol\\
2009& 	45& 	 Henzi SP& 	 Cyclicity in the structure of female baboon social networks& 	 behav ecol sociobiol\\
2009& 	45& 	 Sih A& 	 Social network theory: new insights and issues for behavioral ecologists& 	 behav ecol sociobiol\\
2009& 	5& 	 Mcdonald DB& 	 Young-boy networks without kin clusters in a lek-mating manakin& 	 behav ecol sociobiol\\
2009& 	6& 	 Pempek TA& 	 College students' social networking experiences on Facebook& 	 j appl dev psychol\\
2009& 	6& 	 Borgatti SP& 	 Network Analysis in the Social Sciences& 	 science\\
2009& 	6& 	 Chen W& 	 Efficient Influence Maximization in Social Networks& 	 book\\
2009& 	6& 	 Clauset A& 	 Power-Law Distributions in Empirical Data& 	 siam rev\\
2009& 	6& 	 Eagle N& 	 Inferring friendship network structure by using mobile phone data& 	 p natl acad sci usa\\
2010& 	1245& 	 Voelkl B& 	 Simulation of information propagation in real-life primate networks: longevity, fecundity, fidelity& 	 behav ecol sociobiol\\
2010& 	145& 	 Franks DW& 	 Sampling animal association networks with the gambit of the group& 	 behav ecol sociobiol\\
2010& 	45& 	 Drewe JA& 	 Who infects whom? Social networks and tuberculosis transmission in wild meerkats& 	 p roy soc b-biol sci\\
2010& 	35& 	 Lea AJ& 	 Heritable victimization and the benefits of agonistic relationships& 	 p natl acad sci usa\\
2010& 	35& 	 Wey TW& 	 Social cohesion in yellow-bellied marmots is established through age and kin structuring& 	 anim behav\\
2010& 	35& 	 Schurch R& 	 The building-up of social relationships: behavioural types, social networks and cooperative breeding in a cichlid& 	 philos t r soc b\\
2010& 	35& 	 Perreault C& 	 A note on reconstructing animal social networks from independent small-group observations& 	 anim behav\\
2010& 	35& 	 Krause J& 	 Personality in the context of social networks& 	 philos t r soc b\\
2010& 	6& 	 Fortunato S& 	 Community detection in graphs& 	 phys rep\\
2010& 	6& 	 Kaplan AM& 	 Users of the world, unite! The challenges and opportunities of Social Media& 	 bus horizons\\
2010& 	6& 	 Centola D& 	 The Spread of Behavior in an Online Social Network Experiment& 	 science\\
2010& 	6& 	 Roblyer MD& 	 Findings on Facebook in higher education: A comparison of college faculty and student uses and perceptions of social networking sites& 	 internet high educ\\
2011& 	1235& 	 Croft DP& 	 Hypothesis testing in animal social networks& 	 trends ecol evol\\
2011& 	1235& 	 Brent LJN& 	 Social Network Analysis in the Study of Nonhuman Primates: A Historical Perspective& 	 am j primatol\\
2011& 	1235& 	 Sueur C& 	 How Can Social Network Analysis Improve the Study of Primate Behavior?& 	 am j primatol\\
2011& 	1235& 	 Lehmann J& 	 Baboon (Papio anubis) Social Complexity-A Network Approach& 	 am j primatol\\
2011& 	1235& 	 Sueur C& 	 How Can Social Network Analysis Improve the Study of Primate Behavior?& 	 am j primatol\\
2011& 	135& 	 Voelkl B& 	 Network Measures for Dyadic Interactions: Stability and Reliability& 	 am j primatol\\
2011& 	1& 	 Clark FE& 	 Space to Choose: Network Analysis of Social Preferences in a Captive Chimpanzee Community, and Implications for Management& 	 am j primatol\\
2011& 	35& 	 Bode NWF& 	 Soc.l netw. and models for collective motion in animals& 	 behav ecol sociobiol\\
2011& 	35& 	 Kanngiesser P& 	 Grooming Network Cohesion and the Role of Individuals in a Captive Chimpanzee Group& 	 am j primatol\\
2011& 	35& 	 Bode NWF& 	 The impact of social networks on animal collective motion& 	 anim behav\\
2011& 	6& 	 Kietzmann JH& 	 Social media? Get serious! Understanding the functional building blocks of social media& 	 bus horizons\\
2011& 	3& 	 Kelley JL& 	 Predation Risk Shapes Social Networks in Fission-Fusion Populations& 	 plos one\\
2012& 	1235& 	 Farine DR& 	 Social network analysis of mixed-species flocks: exploring the structure and evolution of interspecific social behaviour& 	 anim behav\\
2012& 	135& 	 Mourier J& 	 Evidence of social communities in a spatially structured network of a free-ranging shark species& 	 anim behav\\
2012& 	135& 	 Cantor M& 	 Disentangling social networks from spatiotemporal dynamics: the temporal structure of a dolphin society& 	 anim behav\\
2012& 	135& 	 Foster EA& 	 Social network correlates of food availability in an endangered population of killer whales, Orcinus orca& 	 anim behav\\
2012& 	35& 	 Blonder B& 	 Temporal dynamics and network analysis& 	 methods ecol evol\\
2013& 	1235& 	 Aplin LM& 	 Individual personalities predict social behaviour in wild networks of great tits (Parus major)& 	 ecol lett\\
2013& 	135& 	 Wilson ADM& 	 Network position: a key component in the characterization of social personality types& 	 behav ecol sociobiol\\
2013& 	135& 	 Hobson EA& 	 An analytical framework for quantifying and testing patterns of temporal dynamics in social networks& 	 anim behav\\
2013& 	35& 	 Farine DR& 	 Animal social network inference and permutations for ecologists in R using asnipe& 	 methods ecol evol\\
2013& 	35& 	 Krause J& 	 Reality mining of animal social systems& 	 trends ecol evol\\
2013& 	35& 	 Kurvers RHJM& 	 Contrasting context dependence of familiarity and kinship in animal social networks& 	 anim behav\\
2013& 	35& 	 Farine DR& 	 Social organisation of thornbill-dominated mixed-species flocks using social network analysis& 	 behav ecol sociobiol\\
2014& 	1235& 	 Farine DR& 	 Measuring phenotypic assortment in animal social networks: weighted associations are more robust than binary edges& 	 anim behav\\
2014& 	1235& 	 Silk MJ& 	 The importance of fission-fusion social group dynamics in birds& 	 ibis\\
2014& 	135& 	 Pinter-Wollman N& 	 The dynamics of animal social networks: analytical, conceptual, and theoretical advances& 	 behav ecol\\
2014& 	135& 	 Castles M& 	 Social networks created with different techniques are not comparable& 	 anim behav\\
2014& 	135& 	 Boogert NJ& 	 Perching but not foraging networks predict the spread of novel foraging skills in starlings& 	 behav process\\
2014& 	35& 	 Boogert NJ& 	 Developmental stress predicts social network position& 	 biol letters\\
2014& 	35& 	 Godfrey SS& 	 A contact-based social network of lizards is defined by low genetic relatedness among strongly connected individuals& 	 anim behav\\
2014& 	3& 	 Shizuka D& 	 Across-year social stability shapes network structure in wintering migrant sparrows& 	 ecol lett\\
2015& 	1235& 	 Farine DR& 	 Constructing, conducting and interpreting animal social network analysis& 	 j anim ecol\\
2015& 	1235& 	 Farine DR& 	 Selection for territory acquisition is modulated by social network structure in a wild songbird& 	 j evolution biol\\
2015& 	1235& 	 Farine DR& 	 The role of social and ecological processes in structuring animal populations: a case study from automated tracking of wild birds& 	 roy soc open sci\\
2015& 	1235& 	 Farine DR& 	 Proximity as a proxy for interactions: issues of scale in social network analysis& 	 anim behav\\
2015& 	135& 	 Adelman JS& 	 Feeder use predicts both acquisition and transmission of a contagious pathogen in a North American songbird& 	 p roy soc b-biol sci\\
2015& 	35& 	 Silk MJ& 	 The consequences of unidentifiable individuals for the analysis of an animal social network& 	 anim behav\\
2015& 	35& 	 Aplin LM& 	 Consistent individual differences in the social phenotypes of wild great tits, Parus major& 	 anim behav\\
2015& 	35& 	 Farine DR& 	 Estimating uncertainty and reliability of social network data using Bayesian inference& 	 roy soc open sci\\
2015& 	35& 	 Firth JA& 	 Experimental manipulation of avian social structure reveals segregation is carried over across contexts& 	 p roy soc b-biol sci\\
2015& 	35& 	 Farine DR& 	 Interspecific social networks promote information transmission in wild songbirds& 	 p roy soc b-biol sci\\
2016& 	1235& 	 Spiegel O& 	 Socially interacting or indifferent neighbours? Randomization of movement paths to tease apart social preference and spatial constraints& 	 methods ecol evol\\
2016& 	1235& 	 Croft DP& 	 Current directions in animal social networks& 	 curr opin behav sci\\
2016& 	1235& 	 Leu ST& 	 Environment modulates population social structure: experimental evidence from replicated social networks of wild lizards& 	 anim behav\\
2016& 	35& 	 Firth JA& 	 Social carry-over effects underpin trans-seasonally linked structure in a wild bird population& 	 ecol lett\\
2016& 	5& 	 Jacoby DMP& 	 Emerging Network-Based Tools in Movement Ecology& 	 trends ecol evol\\
2017& 	1235& 	 Fisher DN& 	 Analysing animal social network dynamics: the potential of stochastic actor-oriented models& 	 j anim ecol\\
2017& 	1235& 	 Silk MJ& 	 Understanding animal social structure: exponential random graph models in animal behaviour research& 	 anim behav\\
2017& 	1235& 	 Fisher DN& 	 Social traits, social networks and evolutionary biology& 	 j evolution biol\\
2017& 	135& 	 Silk MJ& 	 The application of statistical network models in disease research& 	 methods ecol evol\\
2017& 	35& 	 Farine DR& 	 A guide to null models for animal social network analysis& 	 methods ecol evol\\
2017& 	5& 	 Formica V& 	 Consistency of animal social networks after disturbance& 	 behav ecol\\
2017& 	5& 	 Mourier J& 	 Does detection range matter for inferring social networks in a benthic shark using acoustic telemetry?& 	 roy soc open sci\\
2017& 	3& 	 Spiegel O& 	 What's your move? Movement as a link between personality and spatial dynamics in animal populations& 	 ecol lett\\
2018& 	1235& 	 Montiglio PO& 	 Social structure modulates the evolutionary consequences of social plasticity: A social network perspective on interacting phenotypes& 	 ecol evol\\
2018& 	135& 	 Dougherty ER& 	 Going through the motions: incorporating movement analyses into disease research& 	 ecol lett\\
2018& 	135& 	 Silk MJ& 	 Contact networks structured by sex underpin sex-specific epidemiology of infection& 	 ecol lett\\
2018& 	135& 	 Farine DR& 	 When to choose dynamic vs. static social network analysis& 	 j anim ecol\\
2018& 	135& 	 Sah P& 	 Disease implications of animal social network structure: A synthesis across social systems& 	 j anim ecol\\
2018& 	35& 	 Spiegel O& 	 Where should we meet? Mapping social network interactions of sleepy lizards shows sex-dependent social network structure& 	 anim behav\\
2018& 	35& 	 Sih A& 	 Integrating social networks, animal personalities, movement ecology and parasites: a framework with examples from a lizard& 	 anim behav\\
2018& 	35& 	 Spiegel O& 	 Where should we meet? Mapping social network interactions of sleepy lizards shows sex-dependent social network structure& 	 anim behav\\
2018& 	35& 	 Sih A& 	 Integrating social networks, animal personalities, movement ecology and parasites: a framework with examples from a lizard& 	 anim behav\\
2018& 	5& 	 Blaszczyk MB& 	 Consistency in social network position over changing environments in a seasonally breeding primate& 	 behav ecol sociobiol\\
2018& 	3& 	 Bani-Yaghoub M& 	 A methodology to quantify the long-term changes in social networks of competing species& 	 ecol model\\
\end{longtable}
\end{landscape}

\end{document}